\documentclass[10pt,journal,final,twocolumn]{IEEEtran}
\usepackage{epsfig,graphicx,subfigure,psfrag,amsmath,cases,bm}
\usepackage{latexsym,amssymb,algorithm,mathtools}
\usepackage{algorithmic}
\usepackage{color}
\usepackage{url}
\usepackage{scrtime}
\usepackage{stfloats}
\usepackage{tablefootnote}
\usepackage{cite}
\usepackage{amsfonts,mathabx}
\usepackage{textcomp}
\usepackage{xcolor,capt-of}
\usepackage{multirow}
\usepackage{amsthm}
\usepackage{subfigure}
\usepackage{makecell}
\usepackage{pifont}
\usepackage{amssymb}
\usepackage{geometry}
\allowdisplaybreaks

\newtheorem{lemma}{Lemma}

\newtheorem{Def}{Definition}

\newtheorem{T-Prob}{Transformed Problem}
\newtheorem{Prop}{Proposition}

\DeclareMathOperator{\Tr}{Tr}

\DeclareMathOperator{\mino}{minimize}
\DeclareMathOperator{\diag}{\mathrm{diag}}

\DeclareMathOperator{\subto}{subject\hspace*{2mm}to}

\newcommand{\QED}{\hfill \ensuremath{\blacksquare}}

\newcommand{\myoverline}[1]{\overline{\overline{#1}}}

\begin{document}

\title{Sensing-aided Near-Field Secure Communications with Mobile Eavesdroppers}

\author{\vspace*{-1mm}
    Yiming Xu, \textit{Graduate Student Member, IEEE,} Mingxuan Zheng, Dongfang Xu, \textit{Member, IEEE,} \\ Shenghui Song, \textit{Senior Member, IEEE,} and Daniel Benevides da Costa, \textit{Senior Member, IEEE}
    \vspace*{-9mm}
\thanks{Yiming Xu, Mingxuan Zheng, Dongfang Xu, and Shenghui Song are with the Hong Kong University of Science and Technology, Hong Kong, China (email: \{yxuds\},\{mzhengan\}@connect.ust.hk, \{eedxu, eeshsong\}@ust.hk)); 
D. B. da Costa is with the Interdisciplinary Research Center for Communication Systems and Sensing (IRC-CSS), Department of Electrical Engineering, King Fahd University of Petroleum $\&$ Minerals (KFUPM), Dhahran 31261, Saudi Arabia (email: danielbcosta@ieee.org). 
}
}

\maketitle
\begin{abstract}
The additional degree of freedom (DoF) in the distance domain of near-field communication offers new opportunities for physical layer security (PLS) design. 
However, existing works mainly consider static eavesdroppers, and the related study with mobile eavesdroppers is still in its infancy due to the difficulty in obtaining the channel state information (CSI) of the eavesdropper.
To this end, we propose to leverage the sensing capability of integrated sensing and communication (ISAC) systems to assist PLS design. To comprehensively study the dynamic behaviors of the system, we propose a Pareto optimization framework, where a multi-objective optimization problem (MOOP) is formulated to simultaneously optimize three key performance metrics: power consumption, number of securely served users, and tracking performance, while guaranteeing the achievable rate of the users with a given leakage rate constraint.
A globally optimal design based on the generalized Bender’s decomposition (GBD) method is proposed to achieve the Pareto optimal solutions. To reduce the computational complexity, we further design a low-complexity algorithm based on zero-forcing (ZF) beamforming and successive convex approximation (SCA). Simulation results validate the effectiveness of the proposed algorithms and reveal the intrinsic trade-offs between the three performance metrics. It is observed that near-field communication offers a favorable beam diffraction effect for PLS, where the energy of the information signal is nulled around the eavesdropper and focused on the users.
\end{abstract}


\section{Introduction}
\par
Due to its inherent broadcast nature, wireless communication is vulnerable to potential eavesdropping. As a complement of classical cryptography-based technique, physical layer security (PLS) takes advantage of the characteristics of wireless channels to provide flexible security services by exploiting beamforming and artificial noise techniques \cite{7470273, 7812773, 9133130}.
As a result, PLS has been widely investigated for far-field communications \cite{7464352}. With the increase of carrier frequency, e.g., millimeter wave and terahertz communication, and size of the antenna array, e.g., massive and extremely large multiple-input multiple-output (MIMO), near-field communication began to play more important roles \cite{9738442}.
The unique property of the near-field channel brings new opportunities and challenges for PLS design. For instance, in the far-field region, beam steering is utilized to achieve secure transmission by exploiting the channel orthogonality in the angular domain \cite{9933849}. As a result, when the eavesdropper is located at an angle close to the user, secure transmission becomes very difficult and even infeasible. However, the spherical wave in the near-field creates new degrees of freedom (DoFs) for PLS design in both the angular and distance domains by beam focusing \cite{10220205, 9860745, 10496996}.
\par
Some engaging results have been achieved for the PLS design in near-field communication \cite{9860861, 10436390, liu2024ris}. Anaya-L\'{o}pez \textit{et al.} \cite{9860861} investigated the use of spatial DoF in extra-large MIMO systems for PLS design. It was shown that the distance DoF can help to achieve a higher secrecy rate and reduce the regions where an eavesdropper makes secure communication infeasible. Zhang \textit{et al.} \cite{10436390} proposed a two-stage algorithm to design the hybrid beamforming for near-field secure transmission. Liu \textit{et al.} \cite{liu2024ris} studied the extremely large-scale reconfigurable intelligent surface (XL-RIS) aided covert communication in the near-field region. The achievable covert rate was maximized by jointly
optimizing the hybrid beamformers at the base station (BS) and the phase shift at the XL-RIS.
\par
However, most existing works assume that the channel state information (CSI) of the eavesdropper is known, which is difficult to achieve in practice because the eavesdropper is uncooperative. One promising solution is to exploit the sensing capability of integrated sensing and communication (ISAC) systems to locate the eavesdropper. Several works utilized sensing to aid secure communication in the far-field region \cite{9199556, 10227884, 10349846}. In particular, Su \textit{et al.} \cite{9199556} investigated the PLS design in MIMO systems where the signal-to-noise ratio (SNR) at the eavesdroppers was minimized while guaranteeing
the signal-to-interference-plus-noise ratio (SINR) requirement of the users. Furthermore, Su \textit{et al.} \cite{10227884} proposed to utilize an omni-directional wave to estimate the angle of departure (AoD) of the eavesdropper, and the result was utilized for secure transmission. Xu \textit{et al.} \cite{10349846} proposed a time-splitting design for sensing-assisted secure communications.
In the first phase, the system estimated the location of the eavesdropper which was used to enhance the PLS in the second phase.
The time allocation and beamforming policy were optimized to maximize the sum rate.
\par
Despite the above progress in sensing-assisted PLS design, the dynamic behavior of the eavesdropper has not been considered mainly due to two reasons. On the one hand, acquiring the CSI of the moving eavesdropper is challenging. On the other hand, the movement of the eavesdropper may cause a dynamic infeasible region for secure communications. This necessitates dynamic user scheduling design,
and the pioneering work of \cite{10345500} considered user scheduling based on the channel correlation coefficients between users and the eavesdropper in far-field communication. However, to the best of the authors’ knowledge, the optimal design for secure communication with mobile eavesdroppers in both far and near fields are still not available in the literature. 
\par
This paper investigates sensing-aided PLS design for near-field communication with mobile eavesdroppers, and the results can also be applied to the far-field scenario. In particular, we consider an ISAC system where the BS transmits dedicated sensing signals toward the eavesdropper for both sensing and information jamming purposes. In each time slot, the system jointly designs the communication signals, dedicated sensing signals, and user scheduling strategy based on the achievable result from the previous time slot. The echoes received by the BS are used to acquire the CSI of the eavesdropper by tracking its angle, distance, and velocity with extended Kalman filter (EKF). 
To fully characterize the system behavior, a Pareto optimization framework is proposed to simultaneously optimize the power consumption, number of securely served users, and tracking performance, with given achievable rate and leakage rate requirements. An optimization framework based on the constraint method and generalized Bender’s decomposition (GBD) is developed to achieve the Pareto optimal solutions.







\par
The main contributions of this paper are summarized as follows.
\begin{itemize}
\item We investigate the PLS design in near-field communication systems with mobile eavesdroppers, where sensing is utilized to track the eavesdropper by utilizing EKF. 
A multi-objective optimization
problem (MOOP) is formulated to simultaneously
optimize the power consumption, the number of users that can be served securely,
and the tracking performance, while guaranteeing the achievable rate requirement of the users and restricting the leakage rate
to the eavesdropper.
\item
To obtain the Pareto optimal solution, we first employ the constraint method to transform the MOOP into a single-objective optimization problem (SOOP). The resulting SOOP is a mixed integer nonlinear programming (MINLP) problem which is NP-hard. Then, we exploit the GBD theory and implement a series of transformations to develop an algorithm that
achieves the global optimum of the resulting optimization problem with guaranteed convergence. Furthermore, to strike a balance between complexity and optimality, we propose a low-complexity design based on zero-forcing (ZF) beamforming and successive convex approximation (SCA).
\item
The effectiveness of the proposed GBD-based method is validated by simulation results. It is observed that near-field PLS presents a beam diffraction effect in which the energy of the information beam is nulled around the eavesdropper and focused on the users. Moreover, the low-complexity design achieves comparable performance as the optimal solution within a few iterations, indicating its suitability for the real-time application.
\end{itemize}

\textit{Notations:} 
Vectors and matrices are denoted by boldface lowercase and boldface capital letters, respectively. $\mathbb{R}^{M\times N}$ and $\mathbb{C}^{M\times N}$ represent the space of the $M\times N$ real-valued and complex-valued matrices, respectively. $|\cdot|$ and $||\cdot||_2$ denote the absolute value of a complex scalar and the $l_2$-norm of a vector, respectively. $\mathbb{H}^N$ denotes the set of complex Hermitian matrices of dimension $N$.
$(\cdot)^T$ and $(\cdot)^H$ stand for the transpose and the conjugate transpose operator, respectively. $\mathbf{I}_{N}$ refers to the $N$ by $N$ identity matrix. $\mathrm{tr}(\mathbf{A})$ and $\mathrm{rank}(\mathbf{A})$ denote the trace and the rank of matrix $\mathbf{A}$, respectively. $\mathbf{A}\succeq\mathbf{0}$ indicates that $\mathbf{A}$ is a positive semidefinite matrix. $\Re\{\cdot\}$ and $\Im\{\cdot\}$ represent the real and imaginary parts of a complex number, respectively. Vectorization of matrix $\mathbf{A}$ is denoted by $\mathrm{vec}(\mathbf{A})$, and $\mathbf{A}\otimes\mathbf{B}$ represents the Kronecker product between two matrices $\mathbf{A}$ and $\mathbf{B}$. $\mathbb{E}[\cdot]$ refers to statistical expectation. $\overset{\Delta }{=}$ and $\sim$ stand for ``defined as'' and ``distributed as'', respectively. $\mathcal{O}(\cdot)$ is the big-O notation.
\begin{figure}[t]
\centering
\includegraphics[width=3.2in]{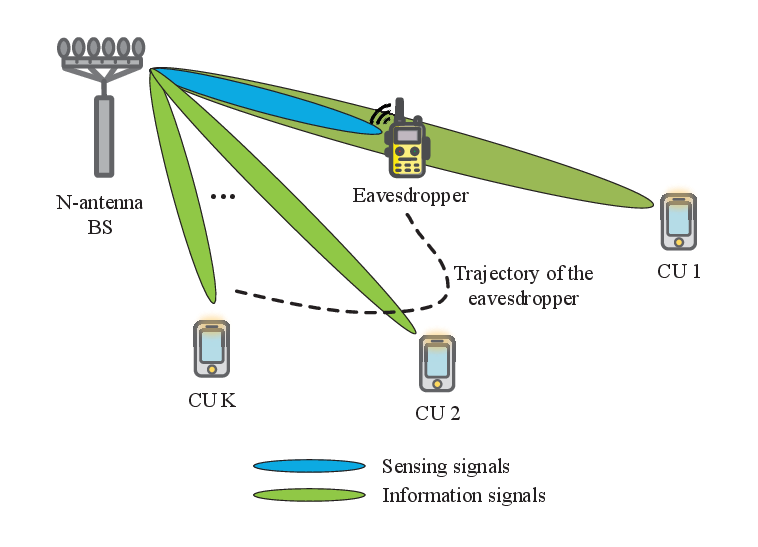}
\vspace*{-6mm}
\caption{Illustration of the considered secure near-field communication system.}
\label{figure:system_model}
\vspace*{0mm}
\end{figure}

\begin{figure}[t]
\centering
\vspace*{-2mm}
\includegraphics[width=3.5in]{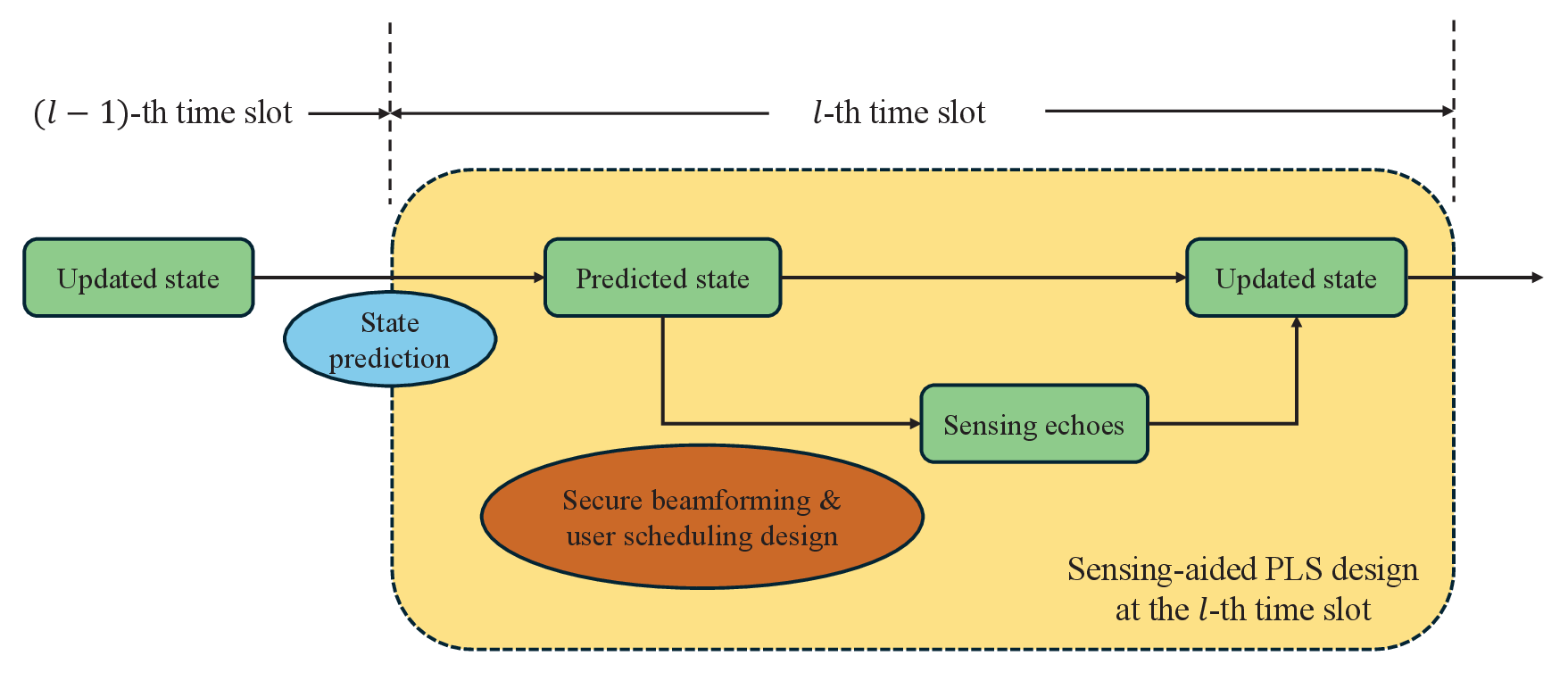}
\vspace*{-4mm}
\caption{The framework of the sensing-aided PLS design.}
\label{figure:protocol}
\vspace*{0mm}
\end{figure}
\section{System Model}
As shown in Fig. \ref{figure:system_model}, we consider 
a near-field communication system where one BS communicates with
$K$ users and one eavesdropper moves around to wiretap the information of the users. The BS is equipped with a uniform linear array (ULA) consisting of $N$ antennas, while the $K$ users and the eavesdropper are equipped with a single antenna.  At the beginning of time slot $l$, the BS acquires the predicted state of the eavesdropper 
based on the estimation results in time slot $l-1$.
With the predicted state of the eavesdropper, the BS performs user scheduling and secure transmission. Based on the predicted state information and the sensing echoes, the BS will update the state information of the eavesdropper
by EKF. The framework of the sensing-aided PLS design is illustrated in Fig. \ref{figure:protocol}. 
\subsection{Signal Model}
We define the set $\mathcal{K}\overset{\Delta }{=}\left\{1,\cdots, K\right\}$ to collect the indices of $K$ users. 
At time slot $l \in \mathcal{L}$, the transmitted signal from the BS can be expressed as
\begin{eqnarray}
\mathbf{x}[l] = \underset{k\in\mathcal{K}}{\sum} e_k[l] \mathbf{w}_k[l] s_k[l] + \mathbf{z}[l],
\end{eqnarray}
where $e_k[l]$ is a binary variable indicating the scheduling status of the $k$-th user at the $l$-th time slot. Specifically, $e_k[l]=1$ indicates that the $k$-th user is scheduled for communication. Otherwise, the $k$-th user is not served. Here, $\mathbf{w}_k[l] \in \mathbb{C}^N$ denotes the beamforming vector for the $k$-th user at the $l$-th time slot and $s_k[l] \sim \mathcal{CN}(0,1)$ is the information symbol for the $k$-th user.
The sensing signal $\mathbf{z}[l] \sim \mathcal{CN}(\mathbf{0}, \mathbf{Z}[l])$ is independent of the communication signals and we denote $\mathbf{Z}[l] \overset{\triangle}{=} \mathbb{E}[\mathbf{z}[l] \mathbf{z}^H[l]] \in \mathbb{H}^N$ as the covariance matrix of the sensing signal \cite{10436719}.
As a result, the covariance matrix of the transmitted signal is given by
\begin{eqnarray}
\mathbf{R}[l] = \underset{k \in \mathcal{K}}{\sum} e_k[l] \mathbf{w}_k[l] \mathbf{w}_k^H[l] + \mathbf{Z}[l].
\end{eqnarray}
\par
With proper synchronization \cite{nasir2016timing}, the received signal of the $k$-th user is given by
\begin{eqnarray} \label{user_signal}
\hspace*{-1mm}y_{k}[l] &\hspace*{-2mm}=\hspace*{-2mm}& \underbrace{e_k[l] \mathbf{h}_k^H[l] \mathbf{w}_k[l] s_k[l]}_{\text{Desired information signal}}\hspace*{1mm}+\hspace*{1mm}\underbrace{\mathbf{h}_k^H[l] \mathbf{z}[l]}_{\text{Interference from sensing signal}}  \notag \\
&\hspace*{-2mm} + \hspace*{-2mm}& \underbrace{\underset{k' \in \mathcal{K} \setminus \{k\} }{\sum} e_{k'}[l] \mathbf{h}_k^H[l] \mathbf{w}_{k'}[l] s_{k'}[l]}_{\text{Multiuser interference}}
\hspace*{1mm}+\hspace*{1mm} n_k[l],
\end{eqnarray}
where $n_k[l] \sim \mathcal{CN}(0,\sigma_{k}^2)$ denotes the additive white Gaussian noise (AWGN) at the $k$-th user. The channel vector between the BS and the $k$-th user $\mathbf{h}_k[l]$ is given by
\begin{eqnarray}
\mathbf{h}_k[l] = \frac{\sqrt{\alpha}}{d_k[l]} \mathbf{a}(\theta_k[l], d_k[l]),
\end{eqnarray}
where $\alpha = ( \frac{\lambda_c}{4 \pi} )^2$ with $\lambda_c$ denoting the wavelength of the carrier. Vector $\mathbf{a}(\theta_k[l], d_k[l])$ is the near-field array response vector where $\theta_k[l]$ and $d_k[l]$ denote the angle and distance of the $k$-th user with respect to the BS at the $l$-th time slot, respectively. Without loss of generality, we set the origin of the coordinates at the center of the ULA where the x-axis aligns with the ULA.
Further denote the antenna index as $n \in \{ -\Tilde{N},\dots, \Tilde{N} \}$ where $N=2 \Tilde{N} + 1$ \cite{10220205}. The $n$-th element of $\mathbf{a}(\theta_k[l], d_k[l])$ is thus given by
\begin{eqnarray} 
&& [\mathbf{a}(\theta_k[l], d_k[l])]_n \notag \\ 
= \hspace{-5mm} && \mathrm{exp}\Big(-j \frac{2 \pi}{\lambda_c}\big(-nd\cos{\theta_k[l]} + \frac{n^2d^2\sin^2{\theta_k[l]}}{2d_k[l]} \big) \Big),
\end{eqnarray}
where $d$ denotes the spacing between adjacent antenna elements.
It follows from \eqref{user_signal} that the achievable rate of the $k$-th user in the $l$-th time slot is given by
\begin{eqnarray}
&&\hspace*{-10mm}R_{\mathrm{info},k}[l] =\notag\\
&&\hspace*{-10mm}\log_2\left(\hspace*{-1mm}1+\hspace*{-0.5mm}\frac{e_k[l] \left| \mathbf{h}^H_k[l] \mathbf{w}_k[l] \right|^2 }{\hspace*{-2mm}\underset{k' \in \mathcal{K}\setminus{k}}{\sum}\hspace*{-1mm}e_{k'}[l] \left| \mathbf{h}_k^H[l]\mathbf{w}_{k'}[l] \right|^2 \hspace*{-1mm}+\hspace*{-0.5mm} \mathbf{h}_k^H[l] \mathbf{Z}[l] \mathbf{h}_k[l] \hspace*{-1mm}+\hspace*{-0.5mm} \sigma_k^2  } \right).
\end{eqnarray}
\par
On the other hand, the received signal at the eavesdropper is given by
\begin{eqnarray}
y_{\mathrm{E}}[l] = \underset{k \in \mathcal{K}}{\sum} e_k[l] \mathbf{h}^H_{\mathrm{E}}[l] \mathbf{w}_k[l]s_k[l]\hspace*{-0.5mm}+\hspace*{-0.5mm}\mathbf{h}^H_{\mathrm{E}}[l] \mathbf{z}[l] + n_{\mathrm{E}}[l],
\end{eqnarray}
where $n_{\mathrm{E}}[l] \sim \mathcal{CN}(0,\sigma_{\mathrm{E}}^2)$ denotes the AWGN. The channel vector between the BS and the eavesdropper is 
given by
\begin{eqnarray}
\mathbf{h}_{\mathrm{E}}[l] = \frac{\sqrt{\alpha}}{d_{\mathrm{E}}[l]} \mathbf{a}(\theta_{\mathrm{E}}[l], d_{\mathrm{E}}[l]),
\end{eqnarray}
where $\theta_{\mathrm{E}}$ and $d_{\mathrm{E}}$ denote the angle and distance of the eavesdropper with respect to the BS, respectively.
In this paper, we consider the worst-case scenario where the eavesdropper 
can cancel the multiuser interference (MUI) \cite{6739367}. As a result, the information leakage rate to the eavesdropper is given by
\begin{eqnarray}
R_{\mathrm{leak},k}[l] = \log_2 \left( 1 + \frac{e_k[l] \left| \mathbf{h}_{\mathrm{E}}^H[l]\mathbf{w}_k[l] \right|^2}{ \mathbf{h}_{\mathrm{E}}^H[l]\mathbf{Z}[l] \mathbf{h}_{\mathrm{E}}[l] + \sigma_{\mathrm{E}}^2} \right).
\end{eqnarray}

\subsection{Radar Measurement Model}
In this subsection, we set up the measurement and state evolution model for the eavesdropper based on EKF.
The received sensing echoes at the BS are given by
\begin{eqnarray} \label{echoes}
\hspace*{-4mm}\mathbf{y}_{\mathrm{s}}(l,t) &\hspace{-3mm} = &\hspace{-3mm} \mathrm{exp}\Big(j2\pi \nu[l] t\Big) \underset{k \in \mathcal{K}}{\sum}e_k[l] \mathbf{H}_{\mathrm{s}}[l] \mathbf{w}_k[l] s_k(l,t-\tau[l]) \notag \\ 
&\hspace{-3mm} + &\hspace{-3mm} \mathrm{exp}\Big(j2\pi \nu[l] t\Big) \mathbf{H}_{\mathrm{s}}[l] \mathbf{z}(l,t-\tau[l]) + \mathbf{n}_{\mathrm{s}}(l,t),
\end{eqnarray}
where $\mathbf{H}_{\mathrm{s}}[l] \overset{\triangle}{=} \frac{\alpha \beta[l]}{d_{\mathrm{E}}^2[l]} \mathbf{a}(\theta_{\mathrm{E}}[l], d_{\mathrm{E}}[l]) \mathbf{a}^H(\theta_{\mathrm{E}}[l], d_{\mathrm{E}}[l])$ denotes the sensing response matrix, variable $\beta[l]$ denotes the radar cross-section of the eavesdropper, which is assumed to follow the Swerling-I target model \cite{10476610}, variables $\tau[l]$ and $\nu[l]$ represent the round-trip delay and Doppler frequency, respectively, and vector $\mathbf{n}_{\mathrm{s}}(l,t)$ denotes the AWGN at the BS.
\par
From \eqref{echoes}, one can obtain the measurements $\mathbf{u}[l]=[\hat{\tau}[l], \hat{\nu}[l], \hat{\theta}_{\mathrm{E}}[l] ]^T$ with $\hat{\tau}[l]$, $\hat{\nu}[l]$ and $\hat{\theta}_{\mathrm{E}}[l]$ denoting the estimated time delay, Doppler frequency, and angle, respectively.
Assume that the clutters, including signals reflected by users, can be effectively suppressed using clutter suppression \cite{skolnik1962introduction}.
Under such circumstances, the measurement model can be expressed as \cite{kay1993fundamentals, 10476610, 9171304}
\begin{eqnarray} \label{measurement_model}
\begin{matrix} \begin{cases}
\hat{\tau}[l] = 2d_{\mathrm{E}}[l]/c + n_{\hat{\tau}}[l], \\
\hat{\nu}[l] = -\frac{2}{\lambda_c}\Big(v_x[l]\cos{\theta_{\mathrm{E}}[l]} + v_y[l] \sin{\theta_{\mathrm{E}}[l]}\Big) + n_{\hat{\nu}}[l], \\
\hat{\theta}_{\mathrm{E}}[l] = \theta_{\mathrm{E}}[l] + n_{\hat{\theta}_{\mathrm{E}}}[l],
\end{cases} \end{matrix}
\end{eqnarray}
where $n_{\hat{\tau}}[l]$, $n_{\hat{\nu}}[l]$, and $n_{\hat{\theta}_{\mathrm{E}}}[l]$ denote the measurement noises, which follow Gaussian distribution with zero mean and variances $\sigma^2_{\hat{\tau}}[l]$, $\sigma^2_{\hat{\nu}}[l]$ and $\sigma^2_{\hat{\theta}_{\mathrm{E}}}[l]$, respectively. The measurement variance is inversely proportional to the SNR given by \cite{9705498}
\begin{eqnarray}
\gamma[l] = \frac{ \alpha^2 \beta^2[l] G N \mathbf{a}^H(\theta_{\mathrm{E}}[l], d_{\mathrm{E}}[l]) \mathbf{Z}[l] \mathbf{a}(\theta_{\mathrm{E}}[l], d_{\mathrm{E}}[l]) }{d_{{\mathrm{E}}}^4[l] \sigma^2_s },
\end{eqnarray}
where $G$ is the number of sensing symbols transmitted during each time slot and $\sigma_{\mathrm{s}}^2$ denotes the noise power at the BS. Then, the measurement variance can be modeled as $\sigma^2_{\hat{\tau}}[l]= a_{\tau}^2/\gamma[l]$, $\sigma^2_{\hat{\nu}}[l]=a_{\nu}^2/\gamma[l]$ and $\sigma^2_{\hat{\theta}_{\mathrm{E}}}[l] = a_{\theta_{\mathrm{E}}}^2 / \gamma[l]$ where $a_{\tau}$, $a_{\nu}$ and $a_{\theta_{\mathrm{E}}}$ are parameters related to the system configurations \cite{9171304}.
\subsection{Tracking Scheme for Eavesdropper}
In this subsection, we present the tracking scheme for the eavesdropper based on EKF. Denote the state parameters of the eavesdropper as $\mathbf{s}=[\theta_{\mathrm{E}}, d_{\mathrm{E}}, v_x, v_y]^T$, where $v_x$, $v_y$ represent the 2D velocity of the eavesdropper.  At the beginning of the $l$-th time slot, the system first obtains the predicted state $\mathbf{s}[l|l-1]$ based on the state evolution model shown in \eqref{state_model} at the top of the next page. Here, $\Delta T$ represents the duration of the time slot, and $n_{\theta_{\mathrm{E}}}[l], n_{d_{\mathrm{E}}}[l], n_{v_x}[l], n_{v_y}[l]$ denote the state evolution noises, which follow Gaussian distribution with zero mean and variances $\sigma_{\theta_{\mathrm{E}}}^2, \sigma_{d_{\mathrm{E}}}^2, \sigma_{v_x}^2$, and  $\sigma_{v_y}^2$, respectively.
\begin{figure*}[t]
	\setcounter{equation}{12}
\begin{eqnarray}
\begin{matrix}
\begin{cases}
\label{state_model}
\theta_{\mathrm{E}}[l|l-1]= \theta_{\mathrm{E}}[l-1]+ \frac{(v_y[l-1] \cos{\theta_{\mathrm{E}}}[l-1] - v_x[l-1] \sin{\theta_{\mathrm{E}}}[l-1])\Delta T}{d_{\mathrm{E}}[l-1]} + n_{\theta_{\mathrm{E}}}[l], \\
d_{\mathrm{E}}[l|l-1]= d_{\mathrm{E}}[l-1]+(v_x[l-1] \cos{\theta_{\mathrm{E}}}[l-1]+v_y[l-1] \sin{\theta_{\mathrm{E}}}[l-1]) \Delta T+n_{d_{\mathrm{E}}}[l],\\
v_x[l|l-1]=v_x[l-1]+n_{v_x}[l], \\
v_y[l|l-1]=v_y[l-1]+n_{v_y}[l].
\end{cases}
\end{matrix}
\end{eqnarray}
\setcounter{equation}{13}
			\hrule
\end{figure*}
\par
For simplicity, we recast the state evolution model and measurement model based on \eqref{measurement_model} and \eqref{state_model} as follows
\begin{eqnarray}
\mathbf{s}[l|l-1] &\hspace*{-2mm}=&\hspace*{-2mm} \mathbf{f}(\mathbf{s}[l-1]) + \mathbf{n}_{\mathrm{e}}[l], \label{predict} \\
\mathbf{u}[l] &\hspace*{-2mm}=& \hspace*{-2mm}\mathbf{g}(\mathbf{s}[l]) + \mathbf{n}_{\mathrm{m}}[l] \label{measurement},
\end{eqnarray}
where functions $f(\cdot)$ and $g(\cdot)$ can be obtained from \eqref{state_model} and \eqref{measurement_model}, respectively. The noise vector of the state evolution and that of the measurement are defined as $\mathbf{n}_{\mathrm{e}}[l] \overset{\triangle}{=} [n_{\theta_{\mathrm{E}}}[l], n_{d_{\mathrm{E}}}[l], n_{v_x}[l], n_{v_y}[l]]^T$ and $\mathbf{n}_{\mathrm{m}}[l] \overset{\triangle}{=} [n_{\hat{\tau}}[l], n_{\hat{\nu}}[l], n_{\hat{\theta}_{\mathrm{E}}}[l]]^T$, respectively. Further denote 
$\mathbf{Q}_{\mathrm{e}} \overset{\triangle}{=} \diag \{ \sigma_{\theta_{\mathrm{E}}}^2, \sigma_{d_{\mathrm{E}}}^2, \sigma_{v_x}^2,  \sigma_{v_y}^2 \}$ and $\mathbf{Q}_{\mathrm{m}}[l] \overset{\triangle}{=} \diag\{ \sigma^2_{\hat{\tau}}[l], \sigma^2_{\hat{\nu}}[l], \sigma^2_{\hat{\theta}_{\mathrm{E}}}[l] \} $ as the covariance matrices for $\mathbf{n}_{\mathrm{e}}[l]$ and $\mathbf{n}_{\mathrm{m}}[l]$, respectively.
\par
The covariance matrix of $\mathbf{s}[l|l-1]$ is given by \cite{kim2018introduction}
\begin{eqnarray}
\mathbf{C}[l|l-1] = \mathbf{F}[l-1] \mathbf{C}[l-1] \mathbf{F}^H[l-1] + \mathbf{Q}_{\mathrm{e}},
\end{eqnarray}
where $\mathbf{F}[l-1] \overset{\triangle}{=} \frac{\partial \mathbf{f}}{\partial \mathbf{s}} \big|_{\mathbf{s}=\mathbf{s}[l-1]}$ and $\mathbf{C}[l-1]$ denotes the covariance matrix of the estimated state at the $(l-1)$-th time slot. Based on the state prediction result, the system determines the beamforming policy and user scheduling strategy. At the same time, sensing echoes are utilized to obtain the measurements $\mathbf{u}[l]$, based on which, the state is updated by
\begin{eqnarray}
\mathbf{s}[l] = \mathbf{s}[l|l-1] + \mathbf{K}[l]\big(\mathbf{u}[l] - \mathbf{g}(\mathbf{s}[l|l-1])\big),
\end{eqnarray}
where $\mathbf{K}[l]$ is the Kalman gain given by
\begin{eqnarray} \label{kalman_gain}
\mathbf{K}[l]\hspace*{-0.5mm}=\hspace*{-0.5mm}\mathbf{C}[l|l\hspace*{-0.4mm}-\hspace*{-0.4mm}1] \mathbf{G}^H[l]\Big(\mathbf{G}[l] \mathbf{C}[l|l\hspace*{-0.4mm}-\hspace*{-0.4mm}1] \mathbf{G}^H[l]\hspace*{-0.5mm}+\hspace*{-0.5mm}\mathbf{Q}_{\mathrm{m}}[l]\Big)^{\hspace*{-0.6mm}-1}\hspace*{-0.5mm},
\end{eqnarray}
with $\mathbf{G}[l] \overset{\triangle}{=} \frac{\partial \mathbf{g}}{\partial \mathbf{s}} \big|_{\mathbf{s}=\mathbf{s}[l|l-1]}$. The corresponding posterior covariance matrix of the estimated state $\mathbf{s}[l]$ is given by
\begin{eqnarray} \label{post_cov}
\mathbf{C}[l] = (\mathbf{I} -\mathbf{K}[l]\mathbf{G}[l] )\mathbf{C}[l|l-1].
\end{eqnarray}
By substituting \eqref{kalman_gain} into \eqref{post_cov}, one can obtain
\begin{eqnarray}
\mathbf{C}[l] = \left( \mathbf{C}^{-1}[l|l-1] + \mathbf{G}^H[l]\mathbf{Q}^{-1}_{\mathrm{e}} \mathbf{G}[l] \right)^{-1}.
\end{eqnarray}
The sensing performance is characterized by the posterior mean squared error (MSE) denoted by $\mathrm{Tr} \left( \mathbf{C}[l] \right)$ \cite{9945983,7069270,10226306}. Next, we deal with the channel error caused by the state estimation error.

\subsection{Channel Error Model}
Since the true value of the channel vector  $\mathbf{h}_{\mathrm{E}}[l]$ is unknown, the system can only utilize the predicted channel  $\mathbf{h}_{\mathrm{E}}[l|l-1] = \frac{\sqrt{\alpha}}{d_{\mathrm{E}}[l|l-1]} \mathbf{a}(\theta_{\mathrm{E}}[l|l-1], d_{\mathrm{E}}[l|l-1])$. 
This inevitably causes prediction errors and necessitates a robust design. 
For that purpose, a tractable model for the prediction error of $\mathbf{a}(\theta_{\mathrm{E}}[l|l-1], d_{\mathrm{E}}[l|l-1])$ with respect to $\theta_{\mathrm{E}}[l|l-1]$, $d_{\mathrm{E}}[l|l-1]$, $\sigma^2_{\theta_{\mathrm{E}}}[l|l-1] \overset{\triangle}{=} \left[\mathbf{C}[l|l-1]\right]_{1,1}$ and $\sigma^2_{d_{\mathrm{E}}}[l|l-1] \overset{\triangle}{=} \left[\mathbf{C}[l|l-1]\right]_{2,2}$ is required.
\par
To this end, we use the widely adopted bounded error model. For ease of illustration, we denote $\Bar{\theta}_{\mathrm{E}} \overset{\triangle}{=} \theta_{\mathrm{E}}[l|l-1]$ and $\Bar{d}_{\mathrm{E}} \overset{\triangle}{=} d_{\mathrm{E}}[l|l-1]$ as the predicted angle and distance, and utilize $\theta_{\mathrm{E}}$ and $d_{\mathrm{E}}$ to represent the true values of angle and distance. Thus, the bounded error model for $\mathbf{a}(\Bar{\theta}_{\mathrm{E}}, \Bar{d}_{\mathrm{E}})$ can be given by \cite{10349846,10153696}
\begin{eqnarray} \label{err_a}
\Delta \mathbf{a}(\bar{\theta}_{\mathrm{E}}, \bar{d}_{\mathrm{E}}, \Delta \theta_{\mathrm{E}}, \Delta d_{\mathrm{E}} ) = \mathbf{a}(\theta_{\mathrm{E}}, d_{\mathrm{E}}) - \mathbf{a}(\bar{\theta}_{\mathrm{E}}, \bar{d}_{\mathrm{E}}),
\end{eqnarray}
where $\Delta \theta_{\mathrm{E}} \overset{\triangle}{=} \theta_{\mathrm{E}} - \Bar{\theta}_{\mathrm{E}} $ and $\Delta d_{\mathrm{E}} = d_{\mathrm{E}} -\Bar{d}_{\mathrm{E}} $ denote the uncertainty of the angle and distance, respectively. According to the three-sigma rule of thumb \cite{10227884}, the uncertainty region of $\Delta \theta_{\mathrm{E}}$ and $\Delta d_{\mathrm{E}}$ are given by 
$\Omega_{\theta_{\mathrm{E}}[l]} \overset{\triangle}{=} \left\{ | \Delta \theta_{\mathrm{E}} | \leq 3 \sigma_{\theta_{\mathrm{E}}}[l|l-1] \right\} $ and $\Omega_{d_{\mathrm{E}}[l]} \overset{\triangle}{=} \left\{ | \Delta d_{\mathrm{E}} | \leq 3 \sigma_{d_{\mathrm{E}}}[l|l-1] \right\}$, respectively.
Then, the norm of $\Delta \mathbf{a}$ can be calculated as
\begin{eqnarray}
&& \left\| \Delta \mathbf{a}(\bar{\theta}_{\mathrm{E}}, \bar{d}_{\mathrm{E}}, \Delta \theta_{\mathrm{E}}, \Delta d_{\mathrm{E}} ) \right\|^2 \notag \\ 
& = & \Big(\mathbf{a}(\theta_{\mathrm{E}}, d_{\mathrm{E}}) - \mathbf{a}(\bar{\theta}_{\mathrm{E}}, \bar{d}_{\mathrm{E}})\Big)^H \Big(\mathbf{a}(\theta_{\mathrm{E}}, d_{\mathrm{E}}) - \mathbf{a}(\bar{\theta}_{\mathrm{E}}, \bar{d}_{\mathrm{E}})\Big) \notag \\
& = & 2N - 2 \varphi(\bar{\theta}_{\mathrm{E}}, \bar{d}_{\mathrm{E}}, \Delta \theta_{\mathrm{E}}, \Delta d_{\mathrm{E}}),
\end{eqnarray}
where $\varphi(\bar{\theta}_{\mathrm{E}}, \bar{d}_{\mathrm{E}}, \Delta \theta_{\mathrm{E}}, \Delta d_{\mathrm{E}})$ is given by \eqref{eq_varphi}, shown at the top of the next page. Note that the estimation error $\Delta \theta_{\mathrm{E}}$ and $\Delta d_{\mathrm{E}}$ are contained in $\varphi$ with a complicated form, making the error bound evaluation very challenging.
To this end, we approximate $\varphi(\bar{\theta}_{\mathrm{E}}, \bar{d}_{\mathrm{E}}, \Delta \theta_{\mathrm{E}}, \Delta d_{\mathrm{E}})$ as follows \cite{10375133}
\setcounter{equation}{23}
\begin{eqnarray}
&& \varphi(\bar{\theta}_{\mathrm{E}}, \bar{d}_{\mathrm{E}}, \Delta \theta_{\mathrm{E}}, \Delta d_{\mathrm{E}}) \approx \sum_{n=-\Tilde{N}}^{\Tilde{N}} \cos \vartheta (n, \bar{\theta}_{\mathrm{E}}, \bar{d}_{\mathrm{E}}, \Delta \theta_{\mathrm{E}}, \Delta d_{\mathrm{E}}) \notag \\
&& \hspace{-4mm} \approx N - \frac{1}{2} \sum_{n=-\Tilde{N}}^{\Tilde{N}} \vartheta^2 (n, \bar{\theta}_{\mathrm{E}}, \bar{d}_{\mathrm{E}}, \Delta \theta_{\mathrm{E}}, \Delta d_{\mathrm{E}}) \notag \\
&& \hspace{-4mm} \overset{\triangle}{=} \hat{\varphi}(\bar{\theta}_{\mathrm{E}}, \bar{d}_{\mathrm{E}}, \Delta \theta_{\mathrm{E}}, \Delta d_{\mathrm{E}}),
\end{eqnarray}
where $\vartheta$ is given by \eqref{emm2}, shown at the top of the next page. The first approximation comes from the first-order Taylor expansion, and the second one comes from the approximation of trigonometric functions.
As a result, the uncertainty region of $\mathbf{a}$ is given by
\setcounter{equation}{25}
\begin{eqnarray}
\Omega_{\mathbf{a}[l]} \overset{\triangle}{=} \left\{ \| \Delta \mathbf{a} \| \leq \sqrt{\beta_{\mathbf{a}}} \right\},
\end{eqnarray}
where 
\begin{eqnarray}
\beta_{\mathbf{a}} \overset{\triangle}{=} \underset{ \substack{\Delta \theta_{\mathrm{E}} \in \Omega_{\theta_{\mathrm{E}}[l]} \\ \Delta d_{\mathrm{E}} \in \Omega_{d_{\mathrm{E}}[l]} } }{\max} 2N - 2\hat{\varphi}(\bar{\theta}_{\mathrm{E}}, \bar{d}_{\mathrm{E}}, \Delta \theta_{\mathrm{E}}, \Delta d_{\mathrm{E}}),
\end{eqnarray}
which is a constant for the robust design below.
\begin{figure*}[t]
	\setcounter{equation}{22}
\begin{eqnarray}
\varphi(\bar{\theta}_{\mathrm{E}}, \bar{d}_{\mathrm{E}}, \Delta \theta_{\mathrm{E}}, \Delta d_{\mathrm{E}}) \hspace*{1mm}=\hspace*{1mm} \sum_{n=-\Tilde{N}}^{\Tilde{N}} \cos \Big( \frac{2\pi}{\lambda_c} \big(nd(\cos \bar{\theta}_{\mathrm{E}} - \cos (\bar{\theta}_{\mathrm{E}} + \Delta \theta_{\mathrm{E}})) + \frac{n^2 d^2}{2}( \frac{\sin^2 (\bar{\theta}_{\mathrm{E}} + \Delta \theta_{\mathrm{E}})}{\bar{d}_{\mathrm{E}} + \Delta d_{\mathrm{E}}} - \frac{\sin^2 \bar{\theta}_{\mathrm{E}}}{\bar{d}_{\mathrm{E}}} ) \big) \Big).
\label{eq_varphi}
\end{eqnarray}
\setcounter{equation}{24}
			\hrule
   \vspace*{2mm}
\setcounter{equation}{24}
\begin{eqnarray} \label{emm2}
&& \vartheta(n, \bar{\theta}_{\mathrm{E}}, \bar{d}_{\mathrm{E}}, \Delta \theta_{\mathrm{E}}, \Delta d_{\mathrm{E}}) \hspace*{1mm}=\hspace*{1mm} \frac{2\pi}{\lambda_c} \Big( \big( nd\sin \bar{\theta}_{\mathrm{E}} + \frac{n^2d^2}{2} \frac{\sin 2\bar{\theta}_{\mathrm{E}}}{\bar{d}_{\mathrm{E}}} \big) \Delta \theta_{\mathrm{E}} - \frac{n^2d^2}{2} \frac{\sin^2 \bar{\theta}_{\mathrm{E}}}{\bar{d}_{\mathrm{E}}^2} \Delta d_{\mathrm{E}} \Big).\label{eq_xi} 
\end{eqnarray}
			\hrule
\end{figure*}
\section{Problem Formulation}
There are three critical performance metrics related to the concerned system: power consumption, number of securely served users, and sensing performance. There exist inherent trade-offs among the three metrics,
which correspond to different system behaviors with varying requirements.
To provide a comprehensive analysis, we formulate a MOOP to simultaneously optimize the three performance metrics while guaranteeing the achievable rate and information leakage rate requirements as follows
\setcounter{equation}{27}
\begin{eqnarray} \label{moop}
&&\hspace{-10mm}\underset{\substack{ \mathbf{w}_k[l], \mathbf{Z}[l], e_k[l] }}{\mino} \hspace*{3mm} \Upsilon(\mathbf{w}_k[l],\mathbf{Z}[l],e_k[l]) \notag\\
&&\hspace*{-8mm}\subto\hspace*{3mm} \mbox{C1:}\hspace*{1mm} R_{\mathrm{info},k}[l] \geq e_k[l] \overline{R}_{\mathrm{info},k}, \hspace*{1mm} \forall k, \notag \\
&&\hspace*{12mm} \mbox{C2:}\hspace*{1mm} \underset{\substack{\Delta \mathbf{a} \in \Omega_{\mathbf{a}[l]} \\ \Delta d_{\mathrm{E}} \in \Omega_{d_{\mathrm{E}}[l]}}}{\max} R_{\mathrm{leak},k}[l] \leq \overline{R}_{\mathrm{leak},k}, \hspace*{1mm} \forall k, \notag \\
&&\hspace*{12mm} \mbox{C3:}\hspace*{1mm} e_k[l] \in \left \{ 0, 1 \right \}, \hspace*{1mm} \forall k.
\end{eqnarray}
The objective function $\Upsilon(\mathbf{w}_k[l],\mathbf{Z}[l],e_k[l])\in\mathbb{R}^{3}$ is given by \
\begin{eqnarray}
&&\hspace*{-12mm}\Upsilon
(\mathbf{w}_k[l],e_k[l],\mathbf{Z}[l])\notag\\
&&\hspace*{-10mm}\overset{\triangle}{=}\left\{ \underset{k \in \mathcal{K}}{\sum} e_k \left\| \mathbf{w}_k[l] \right\|^2\hspace*{-1mm}+\hspace*{-0.5mm}\mathrm{Tr}(\mathbf{Z}[l]),\hspace*{1mm} -\underset{k \in \mathcal{K}}{\sum} e_k[l],\hspace*{1mm}\Tr( \mathbf{C}[l]) \right\}.
\end{eqnarray}
Constraint C1 imposes the achievable rate requirement of each user, i.e., $\overline{R}_{\mathrm{info},k}$, constraint C2 limits the information leakage rate of the served users to be less than the threshold $\overline{R}_{\mathrm{leak},k}$, and constraint C3 indicates the binary property of the user scheduling variables.
\par
The optimality of MOOP is referred to as the Pareto optimality, which is detailed in the following definitions.
\begin{Def}
The achievable performance region is defined as
\begin{eqnarray}
\mathcal{R} =  \left\{ \Upsilon (\mathbf{w}_k[l],\mathbf{Z}[l],e_k[l]) : \forall \left\{ \mathbf{w}_k[l], \mathbf{Z}[l], e_k[l] \right\} \in \mathcal{F} \right\},
\end{eqnarray}
where $\mathcal{F}$ is the feasible solution set defined as
\begin{eqnarray}
\mathcal{F} = \left\{ \left( \mathbf{w}_k[l], \mathbf{Z}[l], e_k[l] \right) : \mbox{C1-C3} \right\}.
\end{eqnarray}
\end{Def}
\par
\begin{Def}
\textit{(Pareto optimal and Pareto boundary):} 
The Pareto boundary $\partial \mathcal{R} \subseteq \mathcal{R}$ consists of all $\mathbf{r} \in \mathcal{R}$ for which there is no $\mathbf{r}' \in \mathcal{R} \setminus \{ \mathbf{r} \}$ with $\mathbf{r}' \geq \mathbf{r}.$ The point $\mathbf{r} \in \partial \mathcal{R}$ is called the Pareto optimal point.
\end{Def}
\par
As widely adopted in the literature, we utilize the Pareto boundary to comprehensively characterize the performance behavior of the considered system. However, the problem in \eqref{moop} is non-convex, which hinders obtaining the optimal solution. Specifically, the non-convexity originates from the fractional term in C1 and C2. Furthermore, user scheduling variables $\{e_k\}$ have a tricky binary form and are coupled with the beamforming policy $\mathbf{w}_k$ and $\mathbf{Z}$. Finally, the problem is intractable due to the semi-infinite constraints in C2 and the complex form of the tracking MSE.
\section{Pareto Optimal Solution of the MOOP}
In this section, we propose an optimization framework to obtain the Pareto optimal solution of the formulated MOOP in \eqref{moop}. To this end, we first use the constrained method \cite{BERUBE200939} to transform \eqref{moop} into a single-objective optimization
problem (SOOP). Then, we develop an algorithm with guaranteed convergence to the global optimum of the SOOP, based on a series of transformations and the GBD method. 
\par
\subsection{Problem Transformation}
According to the constrained method, \eqref{moop} can be transformed into the following SOOP by restricting $\underset{k \in \mathcal{K}}{\sum} e_k$ and $\Tr( \mathbf{C})$ with thresholds $\Gamma_1$ and $\Gamma_2$, respectively.
\begin{eqnarray} \label{constrained_moop}
&&\hspace{-10mm}\underset{\substack{ \mathbf{w}_k[l], \mathbf{Z}[l], e_k[l] }}{\mino} \hspace*{4mm} \underset{k \in \mathcal{K}}{\sum} e_k[l] \left\| \mathbf{w}_k[l] \right\|^2 + \mathrm{Tr}(\mathbf{Z}[l]) \notag\\
&&\hspace*{-8mm}\subto\hspace*{4mm}  \mbox{C1:}\hspace*{1mm} R_{\mathrm{info},k}[l] \geq e_k[l] \overline{R}_{\mathrm{info},k}, \hspace*{1mm} \forall k, \notag \\
&&\hspace*{13mm} \mbox{C2:}\hspace*{1mm} \underset{\substack{\Delta \mathbf{a} \in \Omega_{\mathbf{a}[l]} \\ \Delta d_{\mathrm{E}} \in \Omega_{d_{\mathrm{E}}[l]}}}{\max}      R_{\mathrm{leak},k}[l] \leq \overline{R}_{\mathrm{leak},k}, \hspace*{1mm} \forall k, \notag \\
&&\hspace*{13mm} \mbox{C3:}\hspace*{1mm} e_k[l] \in \left \{ 0, 1 \right \}, \hspace*{1mm} \forall k, \notag \\
&&\hspace*{13mm} \mbox{C4:}\hspace*{1mm} \underset{k \in \mathcal{K}}{\sum} e_k[l] \geq \Gamma_1, \notag \\
&&\hspace*{13mm} \mbox{C5:}\hspace*{1mm} \Tr( \mathbf{C}[l]) \leq \Gamma_2.
\end{eqnarray}
Note that solving problem \eqref{constrained_moop} with all possible values of $\Gamma_1$ and $\Gamma_2$, one can obtain all Pareto optimal solutions on the Pareto boundary $\mathcal{R}$ \cite{miettinen2008introduction}.
Next, we transform \eqref{constrained_moop} into a more tractable form by dealing with the constraints C1, C2, and C5, as well as the variable coupling problem. In the following, the time index $l$ is omitted for simplicity.
\par
First, we define $\mathbf{W}_k \overset{\triangle}{=} \mathbf{w}_k \mathbf{w}_k^H $, $\mathbf{H}_k \overset{\triangle}{=} \mathbf{h}_k \mathbf{h}_k^H$. To recover $\mathbf{w}_k$ from $\mathbf{W}_k$, we further introduce the rank-one constraint $\mbox{C6:} \hspace{1mm} \mathrm{Rank}(\mathbf{W}_k) \leq 1$.
The constraint C1 can then be reformulated as
\begin{eqnarray}
\mbox{C1:}\hspace*{1mm}\frac{e_k \mathrm{Tr} (\mathbf{H}_k \mathbf{W}_k) }{\hspace{-2mm}\underset{k' \in \mathcal{K}\setminus\{k\}}{\sum} \hspace{-1mm} e_{k'} \mathrm{Tr}  (\mathbf{H}_k\mathbf{W}_{k'}) +  \mathrm{Tr}(\mathbf{H}_k \mathbf{Z})  + \sigma_k^2  } \geq \upsilon(e_k),\hspace*{1mm}\forall k.
\end{eqnarray}
where $\upsilon(e_k)\overset{\triangle}{=} 2^{e_k \overline{R}_{\mathrm{info},k}} - 1$. Note that the binary variable $e_k$ appears in the exponent which hinders the obtaining of the optimal solution. To this end, we equivalently substitute $\upsilon(e_k) \overset{\triangle}{=} 2^{e_k \overline{R}_{\mathrm{info},k}} - 1$ with $\overline{\upsilon}(e_k) \overset{\triangle}{=} e_k(2^{\overline{R}_{\mathrm{info},k}} -1)$ by exploiting the binary property of $e_k$. These two functions achieve the same value when $e_k$ takes value from $\{0,1\}$, i.e., $\upsilon(0)=\overline{\upsilon}(0)=0$ and $\upsilon(1)=\overline{\upsilon }(1)=2^{\overline{R}_{\mathrm{info},k}} - 1$.
Next, we eliminate the fractional form of C1 to obtain
\begin{eqnarray}
\overline{\mbox{C1}}\mbox{:} \hspace{-4mm} && e_k \mathrm{Tr} (\mathbf{H}_k \mathbf{W}_k) - \Tilde{R}_{\mathrm{info},k} \underset{k' \in \mathcal{K}\setminus\{k\}}{\sum} \hspace{-2mm} e_k e_{k'} \mathrm{Tr}  (\mathbf{H}_k\mathbf{W}_{k'}) \notag \\ 
&& - \Tilde{R}_{\mathrm{info},k} e_k\mathrm{Tr}(\mathbf{H}_k \mathbf{Z})  - \Tilde{R}_{\mathrm{info},k} \sigma_k^2 e_k  \geq 0, \hspace*{1mm}\forall k,
\end{eqnarray}
where $\Tilde{R}_{\mathrm{info},k} = 2^{\overline{R}_{\mathrm{info},k}} -1$. 
\par
Then, we deal with the semi-infinite constraint C2 which is reformulated as
\begin{eqnarray}
\mbox{C2:} \hspace*{1mm}\underset{\substack{\Delta \mathbf{a} \in \Omega_{\mathbf{a}} \\ \Delta d \in \Omega_{d}}}{\max} \hspace*{1mm}\frac{ \frac{\alpha}{d^2}e_k\mathrm{Tr}\left(\mathbf{a}(\theta, d)\mathbf{a}^H(\theta, d) \mathbf{W}_k\right) }{ \frac{\alpha}{d^2}\mathrm{Tr}\left(\mathbf{a}(\theta, d)\mathbf{a}^H(\theta, d) \mathbf{Z}\right) + \sigma_{\mathrm{E}}^2 } \leq \Tilde{R}_{\mathrm{leak},k}, \hspace*{1mm}\forall k,
\end{eqnarray}
where $\Tilde{R}_{\mathrm{leak},k} = 2^{\overline{R}_{\mathrm{leak},k}} -1$. By eliminating the fractional form, C2 is transformed into
\begin{eqnarray}
\mbox{C2:} \hspace{-6mm} && \underset{\substack{\Delta \mathbf{a} \in \Omega_{\mathbf{a}} \\ \Delta d \in \Omega_{d}}}{\max} e_k \mathrm{Tr}\left(\mathbf{a}(\theta, d)\mathbf{a}^H(\theta, d) \mathbf{W}_k\right) - \Tilde{R}_{\mathrm{leak},k} \sigma_{\mathrm{E}}^2 \frac{d^2}{\alpha}  \notag \\ && - \Tilde{R}_{\mathrm{leak},k} \mathrm{Tr}\left(\mathbf{a}(\theta, d)\mathbf{a}^H(\theta, d) \mathbf{Z}\right) \leq 0,\hspace*{1mm} \forall k.
\end{eqnarray}
Then, we introduce the auxiliary variable $\eta_k \geq 0$ and decompose the constraint C2 into the following two constraints 
\begin{eqnarray}
\mbox{C2a:} \hspace{-6mm} && \underset{\substack{\Delta \mathbf{a} \in \Omega_{\mathbf{a}} }}{\max} e_k \mathrm{Tr}\left(\mathbf{a}(\theta, d)\mathbf{a}^H(\theta, d) \mathbf{W}_k\right) \notag\\ 
&& - \Tilde{R}_{\mathrm{leak},k} \mathrm{Tr}\left(\mathbf{a}(\theta, d)\mathbf{a}^H(\theta, d) \mathbf{Z}\right) \leq \eta_k, \hspace{1mm} \forall k, \label{C2a} \\
\mbox{C2b:} \hspace{-6mm} && \underset{\substack{\Delta d \in \Omega_{d}}}{\min} \Tilde{R}_{\mathrm{leak},k} \sigma_{\mathrm{E}}^2 d^2 - \alpha \eta_k \geq 0, \hspace{1mm} \forall k. \label{C2b}
\end{eqnarray}
Constraints C2a and C2b involve infinitely many constraints because the set $\Omega_{\mathbf{a}}$ and $\Omega_d$ are continuous. To handle this issue, we introduce the following lemma.
\begin{lemma}
\textit{(S-Procedure Lemma \cite{boyd2004convex})} Define two functions $f_i(\mathbf{t}): \mathbb{C}^{N\times 1}\to \mathbb{R}$, $i\in \left \{ 1,2 \right \}$ as
\begin{equation}
f_i(\mathbf{t})= \mathbf{t}^H\mathbf{A}_i\mathbf{t}+2\Re\left \{\mathbf{b}^H_i\mathbf{t}  \right \} + c_i,
\end{equation}
where $\mathbf{A}_i\in \mathbb{H}^N$, $\mathbf{b}_i\in \mathbb{C}^{N\times 1}$, and $\mathrm{c}_i\in \mathbb{R}$. Then, the implication $f_1(\mathbf{t})\leq0 \Rightarrow f_2(\mathbf{t})\leq0$ holds if and only if there exists a variable $\kappa \geq 0$ such that
\begin{equation}
\kappa 
\begin{bmatrix}
\mathbf{A}_1 &  \mathbf{b}_1\\
\mathbf{b}_1^H &  \mathit{c}_1
\end{bmatrix}-\begin{bmatrix}
\mathbf{A}_2 &  \mathbf{b}_2\\
\mathbf{b}_2^H &  \mathit{c}_2
\end{bmatrix}\succeq \mathbf{0}.
\end{equation}
\end{lemma}
Then, by substituting \eqref{err_a} into \eqref{C2a} and \eqref{C2b}, one can rewrite constraints C2a and C2b as follows
\begin{eqnarray}
\hspace{-12mm} && \mbox{C2a:} \hspace{-0.6mm} \underset{\Delta \mathbf{a} \in \Omega_{\mathbf{a}}}{\max} \Delta \mathbf{a}^H \mathbf{S}_k \Delta \mathbf{a} \hspace{-0.6mm} + \hspace{-0.6mm} 2\Re\left\{ \bar{\mathbf{a}}^H \mathbf{S}_k \Delta \mathbf{a} \right\} \hspace{-0.6mm} + \hspace{-0.6mm} \Bar{\mathbf{a}}^H \mathbf{S}_k \Bar{\mathbf{a}} \leq \eta_k, \forall k,\\
\hspace{-12mm} && \mbox{C2b:} \hspace{-0.6mm} \underset{\Delta d \in \Omega_d}{\max} - \Delta d^2 - 2 \Bar{d} \Delta d -\Bar{d}^2 + \frac{\alpha \eta_k}  {\Tilde{R}_{\mathrm{leak},k} \sigma_{\mathrm{E}}^2} \leq 0, \forall k,
\end{eqnarray}
where $\mathbf{S}_k \overset{\triangle}{=} e_k \mathbf{W}_k - \Tilde{R}_{\mathrm{leak},k} \mathbf{Z}$. According to $\textbf{Lemma 1}$, the following implication
\begin{eqnarray}
\Delta \mathbf{a}^H \mathbf{a} \leq \beta_{\mathbf{a}} \Rightarrow \mbox{C2a}
\end{eqnarray}
holds if and only if there exists $\kappa_{1,k} \geq 0$ satisfying
\begin{eqnarray}
\overline{\mbox{C2a}}\mbox{:} 
\begin{bmatrix}
\kappa_{1,k} \mathbf{I}_N & \mathbf{0} \\
\mathbf{0} & - \kappa_{1,k} \beta_{\mathbf{a}} + \eta_k
\end{bmatrix} - \mathbf{U}^H\mathbf{S}_k\mathbf{U} \succeq \mathbf{0}, \hspace{1mm} \forall k,
\end{eqnarray}
where $\mathbf{U} = \begin{bmatrix}
    \mathbf{I}_N & \Bar{\mathbf{a}}
\end{bmatrix} \in \mathbb{C}^{N \times 2}$. Similarly, C2b can be transformed into 
\begin{eqnarray}
\overline{\mbox{C2b}}\mbox{:}
\begin{bmatrix}
\kappa_{2,k} + 1 & \bar{d} \\
\Bar{d} & -\kappa_{2,k} \beta_d + {\bar{d}}^2 - \frac{\alpha \eta_k}{\Tilde{R}_{\mathrm{leak},k} \sigma^2_{\mathrm{E}}}
\end{bmatrix} \succeq \mathbf{0}, \hspace{1mm} \forall k, \label{C2b_}
\end{eqnarray}
where $\kappa_{2,k} \geq 0$ is an auxiliary variable, and $\beta_d \overset{\triangle}{=} 9 \sigma_{d_{\mathrm{E}}}^2$ according to the definition of $\Omega_{d_{\mathrm{E}}[l]}$ in Section II-D.
\par
Next, to deal with the complicated form of the tracking covariance matrix in constraint C5, we introduce auxiliary variables $\zeta_m \geq 0, m \in \mathcal{M} \overset{\triangle}{=} \{1,2,3,4\}$ and transform C5 into the following two constraints
\begin{eqnarray} \label{C5_1}
&& \mbox{C5a}\mbox{:} \hspace{1mm} \left[ \mathbf{C} \right]_{m,m} \leq \zeta_m, \hspace{4mm} \mbox{C5b}\mbox{:} \hspace{1mm} \underset{m \in \mathcal{M}}{\sum} \zeta_m \leq \Gamma_2.
\end{eqnarray}
Then, by employing Schur complement \cite{zhang2006schur}, C5a is equivalently transformed to
\begin{eqnarray} \label{C5_2}
\overline{\mbox{C5a}} \mbox{:} \hspace{1mm}
\begin{bmatrix}
\mathbf{C}^{-1} & \mathbf{q}_m \\
\mathbf{q}_m^T & \zeta_m
\end{bmatrix}
\succeq \mathbf{0}, \forall m,
\end{eqnarray}
where $\mathbf{q}_m \in \mathbb{R}^{4}$ denotes a unit vector whose $m$-th element is 1.
Now, problem \eqref{constrained_moop} is equivalently transformed into 
\begin{eqnarray} \label{opt1}
&&\hspace{-10mm}\underset{\substack{ \mathbf{W}_k, \mathbf{Z}, e_k \\ \kappa_{1,k}, \kappa_{2,k}, \eta_k, \zeta_m \geq 0}}{\mino} \hspace*{3mm}\underset{k \in \mathcal{K}}{\sum} e_k \mathrm{Tr}(\mathbf{W}_k) + \mathrm{Tr}(\mathbf{Z}) \notag\\
&&\hspace*{-6mm}\subto\hspace*{6mm} \overline{\mbox{\mbox{C1}}}, \overline{\mbox{C2a}}, \overline{\mbox{C2b}}, \mbox{C3}, \notag \\
&& \hspace{17mm} \mbox{C4}, \overline{\mbox{C5a}}, \mbox{C5b}, \mbox{C6}.
\end{eqnarray}
However, problem \eqref{opt1} is still non-convex and very challenging to be solved optimally. The difficulty comes from the coupling between the binary variable and the beamforming matrix, the binary constraint on variable $e_k$, and the rank-one constraint C6.
To decouple the binary variable $e_k$ with the beamforming matrix $\mathbf{W}_k$ and $\mathbf{Z}$, we introduce new optimization variables $\overline{\mathbf{W}}_k \overset{\triangle}{=} e_k \mathbf{W}_k$ and $\overline{\mathbf{Z}}_k \overset{\triangle}{=} e_k \mathbf{Z}$ and the following constraints
\begin{eqnarray}
&& \hspace*{-12mm} \mbox{C7a:}\hspace*{1mm}\overline{\mathbf{W}}_{k} \preceq e_k  P_{\mathrm{max}} \mathbf{I},\hspace*{1mm}\forall k,\\
&& \hspace*{-12mm} \mbox{C7b:}\hspace*{1mm}\overline{\mathbf{W}}_{k} \succeq \mathbf{W}_{k}-(1-e_k) P_{\mathrm{max}} \mathbf{I},\hspace*{1mm}\forall k,\\
&& \hspace*{-12mm} \mbox{C7c:}\hspace*{1mm}\overline{\mathbf{W}}_{k} \preceq \mathbf{W}_{k},\hspace*{1mm}\forall k, \hspace{1mm} \mbox{C7d:}\hspace*{1mm}\overline{\mathbf{W}}_{k} \succeq \mathbf{0},\hspace*{1mm}\forall k, \\
&& \hspace*{-12mm} \mbox{C8a:}\hspace*{1mm}\overline{\mathbf{Z}}_{k} \preceq e_k  P_{\mathrm{max}} \mathbf{I},\hspace*{1mm}\forall k,\\
&& \hspace*{-12mm} \mbox{C8b:}\hspace*{1mm}\overline{\mathbf{Z}}_{k} \succeq \mathbf{Z}-(1-e_k) P_{\mathrm{max}} \mathbf{I},\hspace*{1mm}\forall k,\\
&& \hspace*{-12mm} \mbox{C8c:}\hspace*{1mm}\overline{\mathbf{Z}}_{k} \preceq \mathbf{Z},\hspace*{1mm}\forall k, \hspace{1mm} \mbox{C8d:}\hspace*{1mm}\overline{\mathbf{Z}}_{k} \succeq \mathbf{0},\hspace*{1mm}\forall k.
\end{eqnarray}
Constraints C7a-C7d and C8a-C8d linearly characterize the relationship of $\overline{\mathbf{W}}_k = e_k \mathbf{W}_k$ and $\overline{\mathbf{Z}}_k = e_k \mathbf{Z}_k$, respectively. Take C7a-C7d for example, if $e_k=1$, constraint $\mbox{C7a}$ always holds because $\mathrm{Tr}\left( \mathbf{W}_{k} \right) \leq P_{\mathrm{max}}$. $\mbox{C7b}$ becomes $\overline{\mathbf{W}}_k \succeq \mathbf{W}_k$. Together with $\mbox{C7c}$, we have $\overline{\mathbf{W}}_k = \mathbf{W}_{k}$. If $a_q = 0$, constraint $\mbox{C7b}$ always holds. $\mbox{C7a}$ becomes $\overline{\mathbf{W}}_k \preceq \mathbf{0}$. Together with $\mbox{C7d}$, we have $\overline{\mathbf{W}}_k = \mathbf{0}$.
\par
Furthermore, the three variables, i.e., $e_k$, $e_{k'}$ and $\mathbf{W}_{k'}$, are coupled together in the second term of constraint $\overline{\mbox{C1}}$. To handle this, we introduce new variable $\overline{\mathbf{W}}_{kk'} \overset{\triangle}{=} e_k e_{k'} \mathbf{W}_{k'}$ with $(k,k') \in \mathcal{K}' \overset{\triangle}{=} \{ (k,k'),k \in \mathcal{K}, k'\in \mathcal{K}, k\neq k' \}$ and the following constraints
\begin{eqnarray}
&& \hspace*{-12mm} \mbox{C9a:}\hspace*{1mm}\overline{\mathbf{W}}_{kk'} \preceq e_k  P_{\mathrm{max}} \mathbf{I},\hspace*{1mm}\forall (k,k') \in \mathcal{K}', \\
&& \hspace*{-12mm} \mbox{C9b:}\hspace*{1mm}\overline{\mathbf{W}}_{kk'} \preceq e_{k'}  P_{\mathrm{max}} \mathbf{I},\hspace*{1mm}\forall (k,k') \in \mathcal{K}',\\
&& \hspace*{-12mm} \mbox{C9c:}\hspace*{1mm}\overline{\mathbf{W}}_{kk'} \succeq \mathbf{W}_{k'}-(2-e_k-e_{k'}) P_{\mathrm{max}} \mathbf{I},\hspace*{1mm}\forall (k,k') \in \mathcal{K}',\\
&& \hspace*{-12mm} \mbox{C9d:}\hspace*{1mm}\overline{\mathbf{W}}_{kk'} \preceq \mathbf{W}_{k'}, \forall (k,k') \in \mathcal{K}', \\
&& \hspace*{-12mm} \mbox{C9e:}\hspace*{1mm}\overline{\mathbf{W}}_{kk'} \succeq \mathbf{0},\hspace*{1mm}\forall (k,k') \in \mathcal{K}'.
\end{eqnarray}
Constraints C9a-C9e are imposed to ensure that $\overline{\mathbf{W}}_{kk'} = \mathbf{W}_{k'}$ holds only when $e_k = e_{k'} = 1$. Otherwise, we 
have $\overline{\mathbf{W}}_{kk'} = \mathbf{0}$. The variables $e_k \mathbf{W}_k$, $e_k \mathbf{Z}_k$, and $e_k e_{k'} \mathbf{W}_{k'}$ are substituted by $\overline{\mathbf{W}}_k$, $\overline{\mathbf{Z}}_k$, and $\overline{\mathbf{W}}_{kk'}$ in \eqref{opt1}. Constraints $\overline{\mbox{C1}}$ and $\overline{\mbox{C2a}}$ are then transformed equivalently into
\begin{eqnarray}
\myoverline{\mbox{C1}}\mbox{:} \hspace{-4mm} && \mathrm{Tr} (\mathbf{H}_k \overline{\mathbf{W}}_k) - \Tilde{R}_{\mathrm{info},k} \underset{k' \in \mathcal{K}\setminus\{k\}}{\sum} \hspace{-2mm} \mathrm{Tr}  (\mathbf{H}_k\overline{\mathbf{W}}_{kk'}) \notag \\ 
&& - \Tilde{R}_{\mathrm{info},k} \mathrm{Tr}(\mathbf{H}_k \overline{\mathbf{Z}}_k)  - \Tilde{R}_{\mathrm{info},k} \sigma_k^2 e_k  \geq 0, \forall k, \\
\myoverline{\mbox{C2a}}\mbox{:} \hspace{-4mm} && 
\begin{bmatrix}
\kappa_{1,k} \mathbf{I}_N & \mathbf{0} \\
\mathbf{0} & - \kappa_{1,k} \beta_{\mathbf{a}} + \eta_k
\end{bmatrix} - \mathbf{U}^H\overline{\mathbf{S}}_k\mathbf{U} \succeq \mathbf{0}, \label{C2a__} \hspace{1mm} \forall k,
\end{eqnarray}
where matrix $\overline{\mathbf{S}}_k$ is given by $\overline{\mathbf{S}}_k =  \overline{\mathbf{W}}_k - \Tilde{R}_{\mathrm{leak},k} \mathbf{Z}$. Similarly, we obtain $\myoverline{\mbox{C5a}}$ from $\overline{\mbox{C5a}}$.
\par
Then, we exploit semidefinite relaxation (SDR) to remove the rank-one constraint C6. The relaxed problem is given by
\begin{eqnarray} \label{opt2}
&&\hspace{-10mm}\underset{\substack{ \mathbf{W}_k, \mathbf{Z}, e_k \\ \overline{\mathbf{W}}_k, \overline{\mathbf{Z}}_k, \overline{\mathbf{W}}_{kk'}, \\ \kappa_{1,k}, \kappa_{2,k}, \eta_k, \zeta_m \geq 0}}{\mino} \hspace*{4mm}\underset{k \in \mathcal{K}}{\sum} \mathrm{Tr}(\overline{\mathbf{W}}_k) + \mathrm{Tr}(\mathbf{Z}) \notag\\
&&\hspace*{-6mm}\subto\hspace*{6mm} \myoverline{\mbox{C1}}, \myoverline{\mbox{C2a}}, \overline{\mbox{C2b}}, \mbox{C3}, \mbox{C4}, \myoverline{\mbox{C5a}}, \mbox{C5b}, \notag \\
&&\hspace*{17mm} \mbox{C7a-C7d}, \mbox{C8a-C8d}, \mbox{C9a-C9e}.
\end{eqnarray}
The tightness of the SDR relaxation is revealed by the following proposition.
\begin{Prop}
For any optimal solution to problem \eqref{opt2}, one can always construct the equivalent optimal solution with the beamforming matrix $\mathbf{W}^*_k$ satisfying rank-one constraint \mbox{C6}, i.e., $\mathrm{Rank}(\mathbf{W}_k^*)\leq 1$.
\end{Prop}
\textit{Proof:} The proof follows similar steps as in \cite[Appendix B]{xu2024sensing} and thus is omitted here.
\QED
\par
Up to now, problem \eqref{constrained_moop} has been transformed into an NP-hard MINLP problem with some unique properties. Specifically, with given binary variables $e_k$, \eqref{opt2} is a convex optimization problem. For given variables other than $e_k$, \eqref{opt2} is a linear programming problem. These special properties make it possible to find the optimal solution.

\subsection{GBD-based Optimal Design}
In this subsection, we develop a GBD-based framework for optimally solving problem \eqref{opt2}.
To begin with, we denote 
$\bm{\Psi}=\{ \mathbf{W}_k, \mathbf{Z}, \overline{\mathbf{W}}_k, \overline{\mathbf{Z}}_k,  \overline{\mathbf{W}}_{kk'}, \kappa_{1,k}, \kappa_{2,k}, \eta_k, \zeta_m \}$
as the collection of the continuous variables and define the set $\mathcal{F}_{\bm{\Psi}}$ to collect the $\bm{\Psi}$ that satisfies constraints only related to continuous variables as
\begin{eqnarray}
\mathcal{F}_{\bm{\Psi}} \hspace{-3mm} &=& \hspace{-3mm} \left\{ \bm{\Psi}: \myoverline{\mbox{C2a}}, \overline{\mbox{C2b}}, \right. \myoverline{\mbox{C5a}}, \mbox{C5b}, \notag \\
&& \hspace{5mm} \left. \mbox{C7c-C7d}, \mbox{C8c-C8d}, \mbox{C9d-C9e} \right\}.
\end{eqnarray}
Similarly, we denote $\mathbf{e}=\{e_k \}$ as the collection of the binary variables and define the set $\mathcal{F}_{\mathbf{e}}$ to collect $\mathbf{e}$ that satisfies constraints only related to binary variables as
\begin{eqnarray}
\mathcal{F}_{\mathbf{e}} = \left\{ \mathbf{e}: \mbox{C3}, \mbox{C4} \right\}.
\end{eqnarray}
Then, problem \eqref{opt2} can be expressed as
\begin{eqnarray} \label{gbd1}
&&\hspace{-10mm}\underset{\substack{ \bm{\Psi}, \mathbf{e} }}{\mino} \hspace*{4mm} \hspace*{1mm} \underset{k \in \mathcal{K}}{\sum} \mathrm{Tr}(\overline{\mathbf{W}}_k) + \mathrm{Tr}(\mathbf{Z}) \notag\\
&&\hspace*{-11mm}\subto\hspace*{4mm} \myoverline{\mbox{C1}}, \mbox{C7a-C7b}, \mbox{C8a-C8b}, \mbox{C9a-C9c}, \notag \\
&& \hspace{9.5mm} \bm{\Psi} \in \mathcal{F}_{\bm{\Psi}}, \hspace{2mm} \mathbf{e} \in \mathcal{F}_{\mathbf{e}}.
\end{eqnarray}
\par
Note that with fixed binary variables $\mathbf{e}^i$, problem \eqref{gbd1} is a convex problem with respect to $\bm{\Psi}$ given by
\begin{eqnarray} \label{primal}
&&\hspace{-8mm}\underset{\substack{ \bm{\Psi} \in \mathcal{F}_{\bm{\Psi}} }}{\mino} \hspace*{4mm} \hspace*{1mm}  \underset{k \in \mathcal{K}}{\sum} \mathrm{Tr}(\overline{\mathbf{W}}_k) + \mathrm{Tr}(\mathbf{Z}) \notag\\
&&\hspace*{-8mm}\subto\hspace*{2mm} \myoverline{\mbox{C1}}\mbox{:}\hspace*{1mm}  \mathrm{Tr} (\mathbf{H}_k \overline{\mathbf{W}}_k) - \Tilde{R}_{\mathrm{info},k} \underset{k' \in \mathcal{K}\setminus\{k\}}{\sum} \hspace{-2mm} \mathrm{Tr}  (\mathbf{H}_k\overline{\mathbf{W}}_{kk'}) \notag \\ 
&& \hspace{18mm} - \Tilde{R}_{\mathrm{info},k} \mathrm{Tr}(\mathbf{H}_k \overline{\mathbf{Z}}_k)  - \Tilde{R}_{\mathrm{info},k} \sigma_k^2 e_k^i  \geq 0, \hspace{1mm}\forall k, \notag \\
&& \hspace{11mm} \mbox{C7a-C7b}, \mbox{C8a-C8b}, \mbox{C9a-C8c},
\end{eqnarray}
which will be referred to as the $\textit{primal problem}$. The primal problem is convex and can be solved by standard convex optimization solvers such as CVX \cite{grant2014cvx}.
\par
Next, we investigate problem \eqref{gbd1} through its projection onto the $\mathbf{e}$-space, which is given by
\begin{eqnarray} \label{gbd2}
&&\hspace{-10mm}\underset{\substack{ \mathbf{e} }}{\mino} \hspace*{4mm} \underset{\substack{ \bm{\Psi} \in \mathcal{F}_{\bm{\Psi}} }}{\inf} \hspace*{1mm}  \underset{k \in \mathcal{K}}{\sum} \mathrm{Tr}(\overline{\mathbf{W}}_k) + \mathrm{Tr}(\mathbf{Z}) \notag\\
&&\hspace*{-10mm}\subto\hspace*{3mm} \myoverline{\mbox{C1}}, \mbox{C7a-b}, \mbox{C8a-b}, \mbox{C9a-c}, \hspace{2mm} \mathbf{e} \in \mathcal{F}_{\mathbf{e}}.
\end{eqnarray}
By defining the set $\mathcal{E}$ as 
\begin{eqnarray}
\mathcal{E} = \left\{ \mathbf{e}: \exists \bm{\Psi} \in \mathcal{F}_{\bm{\Psi}}, \myoverline{\mbox{C1}}, \mbox{C7a}, \mbox{C7b}, \mbox{C8a}, \mbox{C8b}, \mbox{C9a-C9c}\right\},
\end{eqnarray}
problem \eqref{gbd2} can be concisely expressed as 
\begin{eqnarray} \label{master1}
&&\hspace{-10mm}\underset{\substack{ \mathbf{e} }}{\mino} \hspace*{4mm} \hspace*{1mm}  g(\mathbf{e}) \notag\\
&&\hspace*{-11mm}\subto\hspace*{3mm} \mathbf{e} \in \mathcal{F}_{\mathbf{e}} \cap \mathcal{E},
\end{eqnarray}
where $g(\mathbf{e})$ returns the optimal value of the primal problem \eqref{primal}.
We refer to \eqref{master1} as the $\textit{master problem}$. The equivalence between \eqref{gbd1} and \eqref{master1} is stated by the following lemma.
\begin{lemma} \label{lemma_equ}
\textit{(1)} If $(\bm{\Psi}^*, \mathbf{e}^*)$ is optimal to problem \eqref{gbd1}, then $\mathbf{e}^*$ is optimal to problem \eqref{master1}.
\par
\textit{(2)} If $\mathbf{e}^*$ is optimal to \eqref{master1} and $\bm{\Psi}^*$ is optimal to the primal problem \eqref{primal} with $\mathbf{e}=\mathbf{e}^*$, then $(\bm{\Psi}^*, \mathbf{e}^*)$ is optimal to \eqref{gbd1}.
\par
\textit{(3)} If the problem \eqref{gbd1} is infeasible or unbounded, then the same holds for \eqref{master1} and vice versa.
\end{lemma}
According to $\textbf{Lemma 2}$, problem \eqref{master1} provides a viable route to find the optimal solution to $\eqref{gbd1}$ by viewing it on the $\mathbf{e}$-space, which involves a convex primal problem \eqref{primal}. The challenge of solving \eqref{master1} originates from the fact that the objective function $g(\mathbf{e})$ and the set $\mathcal{E}$ are only known in an implicit way. This difficulty can be tackled by exploiting the dual representation of $g(\mathbf{e})$ and $\mathcal{E}$. Next, we first introduce two types of Lagrangian functions corresponding to the primal problem $\eqref{primal}$ which will be used for deriving the dual representation of $g(\mathbf{e})$ and $\mathcal{E}$.
\par
For any fixed binary variables $\mathbf{e}^i$, there are two cases when solving the primal problem \eqref{primal}, i.e., feasible or infeasible. If the primal problem \eqref{primal} is feasible for given $\mathbf{e}^i$, the corresponding partial Lagrangian function is given by
\begin{eqnarray}
\mathcal{L}(\bm{\Psi}, \mathbf{e}^i, \bm{\Pi})\hspace*{-0.7mm}=\hspace*{-1.5mm}\underset{k \in \mathcal{K}}{\sum} \hspace*{-0.5mm}\mathrm{Tr}(\overline{\mathbf{W}}_k)\hspace*{-0.7mm}+ \hspace*{-0.7mm}\mathrm{Tr}(\mathbf{Z}) \hspace*{-0.7mm}+\hspace*{-0.7mm} h_1(\bm{\Psi}, \bm{\Pi})\hspace*{-0.7mm}+\hspace*{-0.7mm}h_2(\mathbf{e}^i, \bm{\Pi}),
\end{eqnarray}
where $\bm{\Pi} \overset{\triangle}{=} \{\lambda_k, \bm{\Lambda}_k\}$ denotes the collection of dual variables. Variable $\lambda_k$ is the dual variable corresponding to $\myoverline{\mbox{C1}}$ and $\bm{\Lambda}_k \overset{\triangle}{=} \{\bm{\Lambda}_{1,k}, \bm{\Lambda}_{2,k}, \bm{\Lambda}_{3,k}, \bm{\Lambda}_{4,k}, \bm{\Lambda}_{5,kk'}, \bm{\Lambda}_{6,kk'}, \bm{\Lambda}_{7,kk'}\}$ contains the dual variables associated with the constraints $\mbox{C7a-b}$, $\mbox{C8a-b}$, and $\mbox{C9a-c}$, respectively. We denote $\mathcal{F}_{\bm{\Pi}}$ as the collection of the dual variables given by 
\begin{eqnarray} \label{dual_variables}
\mathcal{F}_{\bm{\Pi}} \overset{\triangle}{=} \{ (\lambda_k, \bm{\Lambda}_k): \hspace{-5mm} && \lambda_k \geq 0, \bm{\Lambda}_{p,k} \succeq \mathbf{0}, \bm{\Lambda}_{q,kk'} \succeq \mathbf{0}, \notag \\
&& p \in \{1,2,3,4\}, q \in \{5,6,7\} \}.
\end{eqnarray}
Moreover, functions $h_1(\bm{\Psi}, \bm{\Lambda})$ and $h_2(\mathbf{e}^i, \bm{\Lambda})$ are given by \eqref{h1} and \eqref{h2}, respectively.
\begin{figure*}[t]
	\setcounter{equation}{71}
\begin{eqnarray}
h_1(\bm{\Psi}, \bm{\Lambda}) &\hspace{-2mm}=&\hspace{-2mm}  \underset{k \in \mathcal{K}}{\sum} \lambda_k \Big( \Tilde{R}_{\mathrm{info},k} \mathrm{Tr}(\mathbf{H}_k \overline{\mathbf{Z}}_k)+\Tilde{R}_{\mathrm{info},k} \underset{k' \in \mathcal{K}\setminus\{k\}}{\sum} \hspace{-2mm} \mathrm{Tr}  (\mathbf{H}_k\overline{\mathbf{W}}_{kk'}) -\mathrm{Tr} (\mathbf{H}_k \overline{\mathbf{W}}_k) \Big)
\notag \\
&\hspace{-2mm}+&\hspace{-2mm} \underset{k \in \mathcal{K}}{\sum} \Big( \mathrm{Tr}(\bm{\Lambda}_{1,k} \overline{\mathbf{W}}_k) + \mathrm{Tr}\big(\bm{\Lambda}_{2,k}(\mathbf{W}_k - \overline{\mathbf{W}}_k)\big) + \mathrm{Tr}(\bm{\Lambda}_{3,k} \overline{\mathbf{Z}}_k)  + \mathrm{Tr}\big(\bm{\Lambda}_{4,k}(\mathbf{Z} - \overline{\mathbf{Z}}_k)\big) \Big) \label{h1} \notag \\
&\hspace{-2mm}+&\hspace{-2mm} \underset{(k,k')\in \mathcal{K}'}{\sum} \Big( \mathrm{Tr}(\bm{\Lambda}_{5,kk'} \overline{\mathbf{W}}_{kk'}) +\mathrm{Tr}(\bm{\Lambda}_{6,kk'} \overline{\mathbf{W}}_{kk'}) + \mathrm{Tr}\big(\bm{\Lambda}_{7,kk'}(\mathbf{W}_{k'} - \overline{\mathbf{W}}_{kk'})\big) \Big),\\[4mm]
\vspace*{2mm}
h_2(\mathbf{e}^i, \bm{\Lambda}) &\hspace{-2mm}=&\hspace{-2mm}  \underset{k \in \mathcal{K}}{\sum} \lambda_k \Tilde{R}_{\mathrm{info},k} \sigma_k^2 e_k^i -\underset{k \in \mathcal{K}}{\sum} P_{\mathrm{max}} \Big( e_k^i \mathrm{Tr}(\bm{\Lambda}_{1,k}) + (1-e_k^i) \mathrm{Tr}(\bm{\Lambda}_{2,k}) + e_k^i \mathrm{Tr}(\bm{\Lambda}_{3,k}) + (1-e_k^i) \mathrm{Tr}(\bm{\Lambda}_{4,k}) \Big) \notag \\
&\hspace{-2mm}+&\hspace{-2mm} \underset{(k,k') \in \mathcal{K}'}{\sum} P_{\mathrm{max}} \Big( -e_k^i \mathrm{Tr}(\bm{\Lambda}_{5,kk'}) -e_{k'}^i \mathrm{Tr}(\bm{\Lambda}_{6,kk'})  - (2-e_k-e_{kk'})\mathrm{Tr}(\bm{\Lambda}_{7,kk'}) \Big). \label{h2}
\end{eqnarray}
\setcounter{equation}{73}
			\hrule
\end{figure*}
\par
If the primal problem \eqref{primal} is infeasible for a fixed $\mathbf{e}^i$, then we solve the following feasibility problem by relaxing constraint C1
\begin{eqnarray} \label{feasibility_check}
&&\hspace{-9mm}\underset{\substack{ \bm{\Psi} \in \mathcal{F}_{\bm{\Psi}}, \chi \geq 0 }}{\mino} \hspace*{4mm} \hspace*{1mm}  
\chi \notag\\
&&\hspace*{-10mm}\subto\hspace*{2mm} \widetilde{\mbox{C1}}\mbox{:} - \mathrm{Tr} (\mathbf{H}_k \overline{\mathbf{W}}_k) + \Tilde{R}_{\mathrm{info},k} \underset{k' \in \mathcal{K}\setminus\{k\}}{\sum} \hspace{-2mm} \mathrm{Tr}  (\mathbf{H}_k\overline{\mathbf{W}}_{kk'}) \notag \\ 
&& \hspace{14mm} + \Tilde{R}_{\mathrm{info},k} \mathrm{Tr}(\mathbf{H}_k \overline{\mathbf{Z}}_k) + \Tilde{R}_{\mathrm{info},k} \sigma_k^2 e_k^i  \leq \chi, \hspace{1mm}\forall k, \notag \\
&& \hspace{9mm} \mbox{C7a}, \mbox{C7b}, \mbox{C8a}, \mbox{C8b}, \mbox{C9a-C9c},
\end{eqnarray}
where $\chi$ is the auxiliary variable utilized to indicate the constraint violation. Problem \eqref{feasibility_check} is convex with respect to $\bm{\Psi}$ and $\chi$, and is always feasible \cite{sahinidis1991convergence}. The partial Lagrangian function of the feasibility problem \eqref{feasibility_check} is given by
\begin{eqnarray}
\overline{\mathcal{L}}(\bm{\Psi}, \mathbf{e}^i, \overline{\bm{\Pi}}) = h_1(\bm{\Psi}, \overline{\bm{\Pi}}) + h_2(\mathbf{e}^i, \overline{\bm{\Pi}}),
\end{eqnarray}
where $\overline{\bm{\Pi}} \overset{\triangle}{=} \{\overline{\lambda}_k, \overline{\bm{\Lambda}}_k\} \in \mathcal{F}_{\overline{\bm{\Pi}}}$ is the set collecting the dual variables associated with constraints $\widetilde{\mbox{C1}}$, $\mbox{C7a}$, $\mbox{C7b}$, $\mbox{C8a}$, $\mbox{C8b}$, and $\mbox{C9a-C9c}$ of problem \eqref{feasibility_check}. Set $\mathcal{F}_{\overline{\bm{\Pi}}}$ is defined in the similar manner as $\mathcal{F}_{\bm{\Pi}}$ in \eqref{dual_variables}.
With the above two types of Lagrangian functions, we construct the dual representation of $g(\mathbf{e})$ and $\mathcal{E}$ to get an explicit formulation of problem \eqref{master1} as stated in the following proposition.
\begin{Prop}
The master problem \eqref{master1} can be equivalently expressed as
\begin{eqnarray} \label{master2}
&&\hspace{-12mm}\underset{\substack{ \mathbf{e} \in \mathcal{F}_{\mathbf{e}}, \mu }}{\mino} \hspace*{4mm} \hspace*{1mm} \mu \notag\\
&&\hspace*{-12mm}\subto\hspace*{1mm} \mbox{C10:} \hspace*{1mm}\hspace{1mm} \xi(\bm{\Psi},\mathbf{e},\bm{\Pi}) \leq \mu,\hspace{1mm} \forall  \bm{\Pi} \in \mathcal{F}_{\bm{\Pi}}, \notag \\
&& \hspace{6mm} \mbox{C11:}\hspace*{1mm} \overline{\xi}(\bm{\Psi},\mathbf{e}, \overline{\bm{\Pi}}) \leq 0,\hspace{1mm} \forall \overline{\bm{\Pi}} \in \mathcal{F}_{\overline{\bm{\Pi}}}, \hspace*{1mm}\underset{k \in \mathcal{K}}{\sum} \overline{\lambda}_k = 1,
\end{eqnarray}
where the support functions $\xi$ and $\overline{\xi}$ are defined as
\begin{eqnarray}
&& \xi(\bm{\Psi},\mathbf{e},\bm{\Pi}) \overset{\triangle}{=} \underset{\bm{\Psi} \in \mathcal{F}_{\bm{\Psi}}}{\min} \mathcal{L}(\bm{\Psi}, \mathbf{e}, \bm{\Pi}), \\
&& \overline{\xi}(\bm{\Psi},\mathbf{e}, \overline{\bm{\Pi}} ) \overset{\triangle}{=} \underset{\bm{\Psi} \in \mathcal{F}_{\bm{\Psi}}}{\min} \overline{\mathcal{L}}(\bm{\Psi}, \mathbf{e}, \overline{\bm{\Pi}}).
\end{eqnarray}
\end{Prop}
\par
\textit{Proof:} Since $\mathcal{F}_{\bm{\Psi}}$ is convex, closed, and bounded, and constraint $\widetilde{\mbox{C1}}$ is convex for fixed $e^i_k \in \mathcal{F}_{\mathbf{e}}$. Then, according to \cite[Theorem 2.2]{geoffrion1972generalized}, 
a point $\mathbf{e} \in \mathcal{F}_{\mathbf{e}}$ also belongs to $\mathcal{E}$ if and only if it satisfies:
\begin{eqnarray} \label{dual_set}
\underset{\substack{\bm{\Psi} \in \mathcal{F}_{\bm{\Psi}} }}{\inf} \hspace{1mm} \overline{\mathcal{L}}(\bm{\Psi}, \mathbf{e}, \overline{\bm{\Pi}}) \leq 0, \hspace{1mm} \forall \hspace{1mm} \overline{\bm{\Pi}} \in \mathcal{F}_{\overline{\bm{\Pi}}}, \underset{k \in \mathcal{K}}{\sum} \overline{\lambda}_k = 1,
\end{eqnarray}
which is the dual representation of $\mathcal{E}$.
Next, we show the dual representation of $g(\mathbf{e})$. Note that the primal problem \eqref{primal} is convex and satisfies Slater's condition. Hence, the strong duality holds. Then, one can express  function $g(\mathbf{e})$ as
\begin{eqnarray} \label{dual_function}
g(\mathbf{e}) = \underset{\bm{\Pi} \in \mathcal{F}_{\bm{\Pi}}}{\sup} \hspace{1mm} \underset{\bm{\Psi} \in \mathcal{F}_{\bm{\Psi}}}{\inf} \mathcal{L}(\bm{\Psi}, \mathbf{e}, \bm{\Pi}), \hspace{1mm} \forall \hspace{1mm} \mathbf{e} \in \mathcal{F}_{\mathbf{e}} \cap \mathcal{E},
\end{eqnarray}
By substituting the dual representation of $\mathcal{E}$ and $g(\mathbf{e})$, i.e., \eqref{dual_set} and \eqref{dual_function}, into the master problem \eqref{master1}, we have
\begin{eqnarray} \label{master3}
&&\hspace{-12mm}\underset{\mathbf{e} \in \mathcal{F}_{\mathbf{e}}}{\mino}\hspace*{4mm} \underset{\bm{\Pi} \in \mathcal{F}_{\bm{\Pi}}}{\sup} \hspace*{2mm} \underset{\bm{\Psi} \in \mathcal{F}_{\bm{\Psi}}}{\inf} \hspace*{2mm} \mathcal{L}(\bm{\Psi}, \mathbf{e}, \bm{\Pi}), \notag \\
&&\hspace*{-12mm}\subto\hspace*{1mm} \underset{\substack{\bm{\Psi} \in \mathcal{F}_{\bm{\Psi}} }}{\inf} \hspace{-1mm} \overline{\mathcal{L}}(\bm{\Psi}, \mathbf{e}, \overline{\bm{\Pi}}) \leq 0, \hspace{1mm} \forall \overline{\bm{\Pi}} \in \mathcal{F}_{\overline{\bm{\Pi}}}, \hspace{1mm}\underset{k \in \mathcal{K}}{\sum} \overline{\lambda}_k = 1.
\end{eqnarray}
By introducing the auxiliary variable $\mu$, \eqref{master3} can be equivalently recast as
\begin{eqnarray} \label{master4}
&&\hspace{-12mm}\underset{\substack{ \mathbf{e} \in \mathcal{F}_{\mathbf{e}}, \mu }}{\mino} \hspace*{4mm} \hspace*{1mm} \mu \notag\\
&&\hspace*{-12mm}\subto\hspace*{0mm} \underset{\bm{\Psi} \in \mathcal{F}_{\bm{\Psi}}}{\inf} \mathcal{L}(\bm{\Psi}, \mathbf{e}, \bm{\Pi}) \leq \mu, \forall \bm{\Pi} \in \mathcal{F}_{\bm{\Pi}}, \notag \\
&& \hspace{5mm} \underset{\substack{\bm{\Psi} \in \mathcal{F}_{\bm{\Psi}} }}{\inf} \hspace{1mm} \overline{\mathcal{L}}(\bm{\Psi}, \mathbf{e}, \overline{\bm{\Pi}}) \leq 0, \hspace{1mm} \forall \overline{\bm{\Pi}} \in \mathcal{F}_{\overline{\bm{\Pi}}}, \underset{k \in \mathcal{K}}{\sum} \overline{\lambda}_k = 1.
\end{eqnarray}
Because $\mathcal{F}_{\bm{\Psi}}$ is convex and compact, the optimal solution of problem \eqref{primal} is bounded. Hence, the infimum can be replaced with a minimum in \eqref{master4} which results in \eqref{master2}. \QED
\par
Note that constraints C10 and C11 involve the inner minimization problems with respect to $\bm{\Psi}$. To handle this, we derive the explicit expression of $\xi$ and $\overline{\xi}$, which is given in the following lemma. 
\begin{lemma}
Denote $\bm{\Psi}^i$ and $\bm{\Pi}^i$ as the primal and dual optimal solutions to problem \eqref{primal} if the primal problem \eqref{primal} is feasible. Denote $\overline{\bm{\Psi}}^i$ and $\overline{\bm{\Pi}}^i$ as the primal and dual optimal solutions to problem \eqref{feasibility_check}.
Then $\xi (\bm{\Psi}, \mathbf{e}, \bm{\Pi})$ and $\overline{\xi}(\bm{\Psi}, \mathbf{e}, \overline{\bm{\Pi}})$, parameterized by $\bm{\Pi}^i$ and $\overline{\bm{\Pi}}^i$, can be calculated as follows, respectively,
\begin{eqnarray}
&&\hspace*{-12mm}\xi(\bm{\Psi},\mathbf{e},\bm{\Pi}^i) \hspace{-1mm}=\hspace{-2mm}\underset{k \in \mathcal{K}}{\sum} \mathrm{Tr}(\overline{\mathbf{W}}_k^i) \hspace{-1mm}+ \hspace{-1mm}\mathrm{Tr}(\mathbf{Z}^i) \hspace{-1mm}+\hspace{-1mm}h_1(\bm{\Psi}^i, \bm{\Pi}^i)\hspace{-1mm}+\hspace{-1mm}h_2(\mathbf{e}, \bm{\Pi}^i). \label{opt_cut} \\
&&\hspace*{-12mm}\overline{\xi}(\bm{\Psi},\mathbf{e}, \overline{\bm{\Pi}}^i ) = h_1(\overline{\bm{\Psi}}^i, \overline{\bm{\Pi}}^i) + h_2(\mathbf{e}, \overline{\bm{\Pi}}^i), \label{fea_cut}
\end{eqnarray}
and the multiplier $\overline{\lambda}_k^i$ contained in $\overline{\bm{\Pi}}^i$ satisfies $\underset{k \in \mathcal{K}}{\sum} \overline{\lambda}_k^i=1$.
\end{lemma}
\par
\textit{Proof:}
First, we derive the expression for $\xi (\bm{\Psi}, \mathbf{e}, \bm{\Pi}^i)$. For feasible \eqref{primal}, according to the optimality condition, we have
\begin{eqnarray} \label{primal_optimality_1}
\hspace{0mm}\bm{\Psi}^i  &\hspace{-3mm}=\hspace{-3mm}& \underset{\bm{\Psi} \in \mathcal{F}_{\bm{\Psi}}}{\arg \min} \hspace{1mm} \mathcal{L}(\bm{\Psi}, \mathbf{e}, \bm{\Pi}^i) \notag \\
&\hspace{-3mm}=\hspace{-3mm}&  \underset{\bm{\Psi} \in \mathcal{F}_{\bm{\Psi}}}{\arg \min} \hspace{1mm} \Big\{\hspace{-0.5mm}\underset{k \in \mathcal{K}}{\sum} \hspace{-0.5mm}\mathrm{Tr}(\overline{\mathbf{W}}_k)\hspace{-0.5mm}+\hspace{-0.5mm}\mathrm{Tr}(\mathbf{Z})\hspace{-0.5mm}+\hspace{-0.5mm}h_1(\bm{\Psi}, \bm{\Pi}^i)\hspace{-0.5mm}+\hspace{-0.5mm}h_2(\mathbf{e}, \bm{\Pi}^i) \Big\} \notag \\
&\hspace{-3mm}=\hspace{-3mm}& \underset{\bm{\Psi} \in \mathcal{F}_{\bm{\Psi}}}{\arg \min} \hspace{1mm} \Big\{\hspace{-0.5mm}\underset{k \in \mathcal{K}}{\sum} \mathrm{Tr}(\overline{\mathbf{W}}_k) + \mathrm{Tr}(\mathbf{Z}) + h_1(\bm{\Psi}, \bm{\Pi}^i) \Big\}.
\end{eqnarray}
Then, the function $\xi$ can be rewritten equivalently as
\begin{eqnarray} \label{primal_optimality_2}
\xi &\hspace{-3mm}=\hspace{-3mm}& \underset{\bm{\Psi} \in \mathcal{F}_{\bm{\Psi}}}{\min} \mathcal{L}(\bm{\Psi}, \mathbf{e}, \bm{\Pi}^i) \notag \\
&\hspace{-3mm}=\hspace{-3mm}& \underset{\bm{\Psi} \in \mathcal{F}_{\bm{\Psi}}}{\min} \left\{ \underset{k \in \mathcal{K}}{\sum} \mathrm{Tr}(\overline{\mathbf{W}}_k) + \mathrm{Tr}(\mathbf{Z}) + h_1(\bm{\Psi}, \bm{\Pi}^i)\right\} + h_2(\mathbf{e}, \bm{\Pi}^i), \notag \\
&\hspace{-3mm}=\hspace{-3mm}&  \underset{k \in \mathcal{K}}{\sum} \mathrm{Tr}(\overline{\mathbf{W}}_k^i) + \mathrm{Tr}(\mathbf{Z}^i) + h_1(\bm{\Psi}^i, \bm{\Pi}^i) + h_2(\mathbf{e}, \bm{\Pi}^i).
\end{eqnarray}
According to the optimality condition of \eqref{feasibility_check}, we have
\begin{eqnarray} \label{feasible_optimality}
\hspace{0mm} ( \overline{\bm{\Psi}}^i, \chi^i )  &\hspace{-3mm}=\hspace{-3mm}& \underset{\bm{\Psi} \in \mathcal{F}_{\bm{\Psi}}, \chi\geq 0}{\arg \min} \hspace{1mm} \mathcal{L}(\bm{\Psi}, \mathbf{e}, \bm{\Pi}^i) + \left( 1-\underset{k \in \mathcal{K}}{\sum} \overline{\lambda}_k^i \right) \chi.
\end{eqnarray}
With the KKT condition 
\begin{eqnarray}
\frac{\partial \left( \overline{\mathcal{L}}(\bm{\Psi}^i, \mathbf{e}, \overline{\bm{\Pi}}^i) + (1-\underset{k \in \mathcal{K}}{\sum} \overline{\lambda}_k^i) \chi \right)}{\partial \chi} = 0,
\end{eqnarray}
we obtain 
$ 1-\underset{k \in \mathcal{K}}{\sum} \overline{\lambda}_k^i = 0$.
Further, \eqref{feasible_optimality} can be written as
\begin{eqnarray}
\hspace{0mm}  \overline{\bm{\Psi}}^i  &\hspace{-3mm}=\hspace{-3mm}& \underset{\bm{\Psi} \in \mathcal{F}_{\bm{\Psi}}}{\arg \min} \hspace{1mm} \mathcal{L}(\bm{\Psi}, \mathbf{e}, \bm{\Pi}^i).
\end{eqnarray}
Then, with the same procedures as \eqref{primal_optimality_1} and \eqref{primal_optimality_2}, we can obtain the explicit expression of $\overline{\xi}$ in \eqref{fea_cut}. \QED

With the explicit expression, the master problem is still intractable due to the infinite number of constraints with respect to the dual variables $\bm{\Psi}$ and $\overline{\bm{\Psi}}$ in C10 and C11. To overcome this problem, an iterative relaxation approach is developed by exploiting the GBD theory. Specifically, in each iteration, 
the master problem \eqref{master2} is relaxed by ignoring a few constraints. We solve the relaxed problem and check whether the solutions satisfy all the ignored constraints. If not, we add the violated constraints to the relaxed master problem and solve the master problem again. The violated constraints are generated by solving the primal problem and the feasibility problem according to \textbf{Lemma 3}.
\par
\subsubsection*{Algorithmic Procedure} 
In the $i$-th iteration, the primal problem \eqref{primal} is solved with fixed $\mathbf{e}^{i-1}$ obtained from the last iteration. If it is feasible, we obtain the corresponding primal and dual optimal variables $\bm{\Psi}^i$ and $\bm{\Pi}^i$, based on which, the optimality cut $\xi (\bm{\Psi}, \mathbf{e}, \bm{\Pi}^i) \leq \mu$ is generated according to \eqref{opt_cut}. If \eqref{primal} is infeasible for the given $\mathbf{e}^{i-1}$, we solve the feasibility problem \eqref{feasibility_check}. The primal and dual optimal solutions $\overline{\bm{\Psi}}^i$ and $\overline{\bm{\Pi}}^i$ are used to generate the feasibility cut $\overline{\xi}(\bm{\Psi}, \mathbf{e}, \overline{\bm{\Pi}}^i) \leq 0$ according to \eqref{fea_cut}. Depending on the feasibility of the primal problem \eqref{primal} with given $e^{(i-1)}$, the optimality cut or the feasibility cut is added to the relaxed master problem given by
\begin{eqnarray} \label{relaxed_master}
&&\hspace{-10mm}\underset{\substack{ \mathbf{e} \in \mathcal{F}_{\mathbf{e}}, \mu }}{\mino} \hspace*{4mm} \hspace*{1mm} \mu \notag\\
&&\hspace*{-10mm}\subto\hspace*{2mm} \mbox{C10:} \hspace{1mm} \xi(\bm{\Psi},\mathbf{e},\bm{\Pi}^p) \leq \mu, \hspace{1mm}\forall p \in \mathcal{I}_{\mathrm{fea}}^i, \notag \\
&& \hspace{9mm} \mbox{C11:} \hspace{1mm}\overline{\xi}(\bm{\Psi},\mathbf{e},\bm{\Pi}^q) \leq 0, \hspace{1mm}\forall q \in \mathcal{I}_{\mathrm{inf}}^i,
\end{eqnarray}
where $\mathcal{I}_{\mathrm{fea}}^i = \left\{ p: \eqref{primal} \text{ is feasible at the } p\text{-th iteration}, p \in \mathcal{I} \right\}$ and $\mathcal{I}_{\mathrm{inf}}^i = \left\{ q: \eqref{primal} \text{ is infeasible at the } q\text{-th iteration}, q \in \mathcal{I} \right\}$, where $\mathcal{I} \overset{\triangle}{=} \left\{ 1,\cdots, i \right\}$ collects the iteration indexes. Note that \eqref{relaxed_master} is a MILP problem and can be optimally solved by standard solvers such as MOSEK \cite{grant2014cvx}.
The solution to the relaxed master problem $\mathbf{e}^{i}$ is substituted into the primal problem to generate the corresponding cuts for the next iteration. Because some constraints are ignored, the optimal value $\mu^i$ gives the lower bound of the original problem. We denote the lower bound as the $LBD$ which is updated by $LBD = \mu^i$.
\par
Since the primal problem is solved with fixed $\mathbf{e}$, the optimal value of the primal problem (if feasible) gives the upper bound of the original problem. The upper bound is denoted as $UBD$ and is updated by $UBD = \underset{i \in \mathcal{I}}{\min} \left\{ g(\mathbf{e}^{i}) \right\}$. The algorithm terminates when the difference between the upper bound and the lower bound is less than the predetermined threshold $\epsilon$, i.e., $UBD-LBD \leq \epsilon$. The algorithm is summarized in $\textbf{Algorithm 1}$. Some remarks on the algorithm are as follows:
\begin{algorithm}[t]
\caption{GBD-based Optimal Design}
\begin{algorithmic}[1]
\small
\STATE Initialize iteration index $i=0$, upper bound $UBD = + \infty$, lower bound $LBD = -\infty$, the feasibility index collection set $\mathcal{I}_{\mathrm{fea}}^0 = \emptyset$, the infeasibility index collection set $\mathcal{I}_{\mathrm{inf}}^0 = \emptyset$, convergence tolerance $0<\epsilon \ll 1$, and randomly generate initial binary variable $\mathbf{e}^0$
\REPEAT 
\STATE Set $i=i+1$
\STATE Solve primal problem \eqref{primal} for given $\mathbf{e}^{i-1}$
\IF{the primal problem \eqref{primal} is feasible}
\STATE Obtain the primal and dual optimal solution $\bm{\Psi}^i$ and $\bm{\Pi}^i$
\STATE Add optimality cut $\xi (\bm{\Psi}, \mathbf{e}, \bm{\Pi}^i) \leq \mu$ to relaxed master problem \eqref{relaxed_master}
\STATE Set $\mathcal{I}_{\mathrm{fea}}^i = \mathcal{I}_{\mathrm{fea}}^{i-1} \cup \{ i \}$ and $\mathcal{I}_{\mathrm{inf}}^i = \mathcal{I}_{\mathrm{inf}}^{i-1}$
\STATE Update the upper bound $UBD = \min \{ UBD, g(\mathbf{e}^{i-1}) \}$
\ELSE
\STATE Solve feasibility problem \eqref{feasibility_check} for given $\mathbf{e}^{i-1}$ and obtain $\overline{\bm{\Psi}}^i$ and $\overline{\bm{\Pi}}^i$
\STATE Add feasibility cut $\overline{\xi}(\bm{\Psi}, \mathbf{e}, \overline{\bm{\Pi}}^i) \leq 0$ to relaxed master problem \eqref{relaxed_master}
\STATE Set $\mathcal{I}_{\mathrm{fea}}^i = \mathcal{I}_{\mathrm{fea}}^{i-1}$ and $\mathcal{I}_{\mathrm{inf}}^i = \mathcal{I}_{\mathrm{inf}}^{i-1} \cup \{ i \}$
\ENDIF
\STATE Solve the relaxed problem \eqref{relaxed_master} and obtain $\mathbf{e}^i$
\STATE Update the lower bound $LBD = \mu^i$
\UNTIL $UBD - LBD \leq \epsilon$
\end{algorithmic}
\end{algorithm}
\par
\textit{i) Convergence and Optimality:} As the algorithm proceeds, more constraints are added to the relaxed master problem. As a result, the feasible region shrinks. Hence, the sequence of the lower bound is non-decreasing. The upper bound sequence is non-increasing because the tightest upper bound is recorded at each iteration. As the algorithm proceeds, we can progressively slash the difference between the upper bound and the lower bound. According to \cite{geoffrion1972generalized}, $\textbf{Algorithm 1}$ terminates in a finite number of iterations for any given $\epsilon \geq 0$ and converges to the global optimal solution of \eqref{gbd1} when $\epsilon \rightarrow 0$.
\par
\textit{ii) Complexity:} For each iteration, the complexity for solving the primal problem \eqref{primal} or the corresponding feasibility problem \eqref{feasibility_check} is $A_1 = K(K+1)N^3 + K^2(K+1)^2N^2 + K^3(K+1)^3$ \cite[Theorem 3.12]{bomze2010interior}. The complexity of solving the relaxed master problem \eqref{relaxed_master} is $A_2 = 2^K$. If the algorithm converges within $B$ iterations, then the total computational complexity of $\textbf{Algorithm 1}$ is $\mathcal{O}(B(A_1+A_2))$. It is worth noting that $\textbf{Algorithm 1}$ does not require enumerating all $2^K$ possible solutions and generally converges within fewer iterations than the exhaustive search which has a complexity of $\mathcal{O}(A_1 A_2)$. 
Moreover, compared with the branch-and-bound method, which is a widely used algorithm for optimal design, GBD decomposes the original problem into two 
easier sub-problems. The dimension of the optimization variables for calculating the corresponding beamforming is smaller in each iteration, leading to less complexity than the branch-and-band method. Note that although solving the MILP \eqref{relaxed_master} by standard solvers also requires the branch-and-bound method, the dimension of the optimization variables in \eqref{relaxed_master} is significantly smaller than that in \eqref{gbd1}. Hence, solving \eqref{relaxed_master} has an affordable complexity.

\section{Low-complexity Design}
The proposed GBD-based optimal design achieves the global optimum but suffers from high computational complexity. To tackle this issue, we adopt the ZF beamforming design \cite{10614246}
\begin{eqnarray}
[\mathbf{w}_1[l],..., \mathbf{w}_K[l], \mathbf{w}_{\mathrm{E}}[l]] = \mathbf{H}[l](\mathbf{H}^H[l] \mathbf{H}[l])^{-1},
\end{eqnarray}
where $\mathbf{H}[l] = [\mathbf{h}_1[l], ..., \mathbf{h}_K[l], \mathbf{h}_{\mathrm{E}}[l]]$. The beamformers for the $k$-th user and the eavesdropper are determined as $e_k[l] \sqrt{p_k[l]} \mathbf{w}_k[l]$ and $\sqrt{p_{\mathrm{E}}[l]} \mathbf{w}_{\mathrm{E}}[l]$, respectively, where $p_k[l]$ and $p_{\mathrm{E}}[l]$ denote the corresponding power allocation. 
The considered MOOP is then expressed as
\begin{eqnarray} \label{low_moop}
&&\hspace{-10mm}\underset{\substack{ p_k[l], p_{\mathrm{E}}[l], e_k[l] }}{\mino} \hspace*{3mm} \overline{\Upsilon}(p_k[l], p_{\mathrm{E}}[l], e_k[l]) \notag\\
&&\hspace*{-8mm}\subto\hspace*{3mm} \mbox{C1:}\hspace*{1mm} R_{\mathrm{info},k}[l] \geq e_k[l] \overline{R}_{\mathrm{info},k}, \hspace*{1mm} \forall k, \notag \\
&&\hspace*{12mm} \mbox{C2:}\hspace*{1mm} \underset{\substack{\Delta \mathbf{a} \in \Omega_{\mathbf{a}[l]} \\ \Delta d_{\mathrm{E}} \in \Omega_{d_{\mathrm{E}}[l]}}}{\max} R_{\mathrm{leak},k}[l] \leq \overline{R}_{\mathrm{leak},k}, \hspace*{1mm} \forall k, \notag \\
&&\hspace*{12mm} \mbox{C3:}\hspace*{1mm} e_k[l] \in \left \{ 0, 1 \right \}, \hspace*{1mm} \forall k.
\end{eqnarray}
where
\begin{eqnarray}
&&\hspace*{-12mm} \overline{\Upsilon}
(p_k[l],p_{\mathrm{E}}[l],e_k[l]) \overset{\triangle}{=} \left\{ \underset{k \in \mathcal{K}}{\sum} e_k[l] p_k[l] \left\| \mathbf{w}_k[l] \right\|^2\hspace*{-1mm} \right. \notag\\ 
&& \left. + p_{\mathrm{E}}[l] \left\| \mathbf{w}_{\mathrm{E}}[l] \right\|^2, \hspace*{1mm} -\underset{k \in \mathcal{K}}{\sum} e_k[l],\hspace*{1mm}\Tr( \mathbf{C}[l]) \right\}.
\end{eqnarray}
Similar to Section IV-A, \eqref{low_moop} is transformed into the following SOOP by employing the constrained method
\begin{eqnarray} \label{low_soop}
&&\hspace{-10mm} \underset{\substack{ p_k[l], p_{\mathrm{E}}[l], e_k[l] }}{\mino} \hspace*{4mm} f_{\mathrm{obj}} \overset{\triangle}{=} \underset{k \in \mathcal{K}}{\sum} e_k[l] p_k[l] \left\| \mathbf{w}_k[l] \right\|^2 + p_{\mathrm{E}}[l] \left\| \mathbf{w}_{\mathrm{E}}[l] \right\|^2 \notag\\
&&\hspace*{-8mm}\subto\hspace*{4mm}  \mbox{C1:}\hspace*{1mm} R_{\mathrm{info},k}[l] \geq e_k[l] \overline{R}_{\mathrm{info},k}, \hspace*{1mm} \forall k, \notag \\
&&\hspace*{13mm} \mbox{C2:}\hspace*{1mm} \underset{\substack{\Delta \mathbf{a} \in \Omega_{\mathbf{a}[l]} \\ \Delta d_{\mathrm{E}} \in \Omega_{d_{\mathrm{E}}[l]}}}{\max}      R_{\mathrm{leak},k}[l] \leq \overline{R}_{\mathrm{leak},k}, \hspace*{1mm} \forall k, \notag \\
&&\hspace*{13mm} \mbox{C3:}\hspace*{1mm} e_k[l] \in \left \{ 0, 1 \right \}, \hspace*{1mm} \forall k, \notag \\
&&\hspace*{13mm} \mbox{C4:}\hspace*{1mm} \underset{k \in \mathcal{K}}{\sum} e_k[l] \geq \Gamma_1, \notag \\
&&\hspace*{13mm} \mbox{C5:}\hspace*{1mm} \Tr( \mathbf{C}[l]) \leq \Gamma_2.
\end{eqnarray}
Note that the dimension of the optimization variables in \eqref{low_soop} is much smaller than that of problem \eqref{constrained_moop}. We avoid directly dealing with the NP-hard MINLP problem by exploring the penalty method and SCA, so that the problem can be solved with polynomial complexity. The time index $l$ is omitted in the following.
\par
First, we introduce variable $\overline{p}_k \overset{\triangle}{=} e_k p_k$. Constraint C1 is equivalently transformed into 
\begin{eqnarray}
\widetilde{\mbox{C1}}\mbox{:} \hspace{1mm} \overline{p}_k \mathrm{Tr}(\mathbf{H}_k \mathbf{W}_k) - \Tilde{R}_{\mathrm{info},k} \sigma_k^2 e_k  \geq 0,\hspace{1mm} \forall k.
\end{eqnarray}
Similar to the transformations for $\myoverline{\mbox{C2a}}$ and $\overline{\mbox{C2b}}$ in \eqref{C2a__} and \eqref{C2b_}, we transform C2 into $\widetilde{\mbox{C2a}}$ and $\widetilde{\mbox{C2b}}$ by introducing $\kappa_{1,k}, \kappa_{2,k}, \eta_k$ and omit the expression of $\widetilde{\mbox{C2a}}$ and $\widetilde{\mbox{C2b}}$ for simplicity. Note that $\overline{\mathbf{S}} = \overline{p}_k \mathbf{W}_k - p_{\mathrm{E}} \Tilde{R}_{\mathrm{leak},k} \mathbf{W}_{\mathrm{E}}$ in $\widetilde{\mbox{C2a}}$ is different from that in $\myoverline{\mbox{C2a}}$. In addition, we handle the constraint C5 by the same transformations as in \eqref{C5_1} and \eqref{C5_2} by introducing $\zeta_m$. Constraint $\mbox{C5}$ is then transformed into $\widetilde{\mbox{C5a}}$ and $\widetilde{\mbox{C5b}}$, which has the same structure as $\overline{\mbox{C5a}}$ and C5b, respectively. Next, we replace the relation $\overline{p}_k=e_k p_k$ with the following equations
\begin{eqnarray}
&& \hspace*{-12mm} \mbox{C12a:}\hspace*{1mm}\overline{p}_k \leq e_k  P_{\mathrm{max}} ,\hspace*{1mm}\forall k,\\
&& \hspace*{-12mm} \mbox{C12b:}\hspace*{1mm}\overline{p}_k \geq p_k-(1-e_k) P_{\mathrm{max}} ,\hspace*{1mm}\forall k,\\
&& \hspace*{-12mm} \mbox{C12c:}\hspace*{1mm}\overline{p}_k \leq p_k,\hspace*{1mm}\forall k, \hspace{1mm} \mbox{C12d:}\hspace*{1mm}\overline{p}_k \geq 0,\hspace*{1mm}\forall k,
\end{eqnarray}
\par
The main difficulty lies in the binary constraint C3. To handle this, we equivalently rewrite C3 as
\begin{eqnarray}
\mbox{C3a:} \hspace{1mm} \underset{k \in \mathcal{K}}{\sum}(e_k - e_k^2) \leq 0, \hspace{2mm} \mbox{C3b:} \hspace{1mm} 0 \leq e_k \leq 1,\hspace{1mm}\forall k.
\end{eqnarray}
Note that constraint C3a is the difference between two convex functions, which is intrinsically non-convex. To circumvent this obstacle, we employ the penalty method and take constraint C3a as a penalty term in the objective function \cite{nocedal1999numerical}. The resulting optimization problem is given by
\begin{eqnarray} \label{plus_penalty}
&&\hspace{-10mm}\underset{\substack{ p_k, \overline{p}_k, p_{\mathrm{E}}, e_k }}{\mino} \hspace*{4mm} \hspace*{1mm} \overline{f}_{\mathrm{obj}} \overset{\triangle}{=} f_{\mathrm{obj}} + \varrho \underset{k\in\mathcal{K}}{\sum}(-e_k^2 + e_k)
\notag\\
&&\hspace*{-10mm}\subto\hspace*{2mm} \widetilde{\mbox{C1}}, \widetilde{\mbox{C2a}}, \widetilde{\mbox{C2b}}, \mbox{C3b}, \mbox{C4}, \notag \\
&&\hspace*{9mm} \widetilde{\mbox{C5a}}, \widetilde{\mbox{C5b}}, \mbox{C12a-C12d},
\end{eqnarray}
where $\varrho \gg 0$ is the penalty factor to enforce constraint C3a hold. Then, we utilize SCA to tackle the non-convex penalty term.
A convex surrogate function for the objective function $\overline{f}_{\mathrm{obj}}$ is constructed by applying the first-order Taylor expansion on the term $-e_k^2$. The algorithm proceeds iteratively and the problem in the $i$-th iteration is given by
\begin{eqnarray} \label{after_SCA}
&&\hspace{-10mm}\underset{\substack{ p_k, \overline{p}_k, p_{\mathrm{E}}, e_k }}{\mino} \hspace*{4mm} \hspace*{1mm} \myoverline{f}_{\mathrm{obj}}^{(i)}
\notag\\
&&\hspace*{-10mm}\subto\hspace*{2mm} \widetilde{\mbox{C1}}, \widetilde{\mbox{C2a}}, \widetilde{\mbox{C2b}}, \mbox{C3b}, \mbox{C4}, \notag \\
&&\hspace*{9mm} \widetilde{\mbox{C5a}}, \widetilde{\mbox{C5b}}, \mbox{C12a-C12d},
\end{eqnarray}
where the objective function is given by
\begin{eqnarray}
\myoverline{f}_{\mathrm{obj}}^{(i)} \overset{\triangle}{=} f_{\mathrm{obj}} + \varrho \underset{k\in\mathcal{K}}{\sum}\left((1-2e_k^{(i-1)})e_k + (e_k^{(i-1)})^2 \right),
\end{eqnarray}
with $e_k^{(i-1)}$ obtained from the $(i-1)$-th iteration. The objective function $\myoverline{f}_{\mathrm{obj}}^{(i)}$ is convex with respect to the optimization variables. Hence, the problem \eqref{after_SCA} is convex and can be solved by standard optimization solver CVX.
\par
The ZF-SCA-based low-complexity design is summarized in \textbf{Algorithm 2}.
The objective function of problem \eqref{plus_penalty} is upper bounded by the minimum of problem \eqref{after_SCA}. Iteratively solving problem \eqref{after_SCA} by \textbf{Algorithm 2} monotonically tightens the upper bound and \textbf{Algorithm 2} is guaranteed to converge in polynomial time \cite{dinh2010local}. Moreover, as $\varrho \rightarrow \infty$, the limit of any convergent sequence of the proposed algorithm is a stationary point of problem \eqref{low_soop} \cite{nocedal1999numerical}.
\par
\par

\begin{algorithm}[t]
\caption{ZF-SCA-based Low-complexity Design}
\begin{algorithmic}[1]
\small
\STATE Set iteration index $i=1$, error tolerance factor $0<\epsilon\ll1$, and penalty factor $\varrho$. Initialize the user scheduling variables $e_k^{(0)} = 0, \forall k$
\REPEAT
\STATE Solve problem \eqref{after_SCA} for fixed $e_k^{(i-1)}, \forall q$
\STATE Update $e_k^{(i)}$
\STATE Set $i=i+1$
\UNTIL $\frac{\left|\overline{f}^{(i)}_{\mathrm{obj}}-f_{\mathrm{obj}}^{(i-1)}\right|}{f_{\mathrm{obj}}^{(i)}}\leq \epsilon$
\end{algorithmic}
\end{algorithm}
\section{Numerical Results}

\begin{figure}[t]
\centering
\includegraphics[width=3.2in]{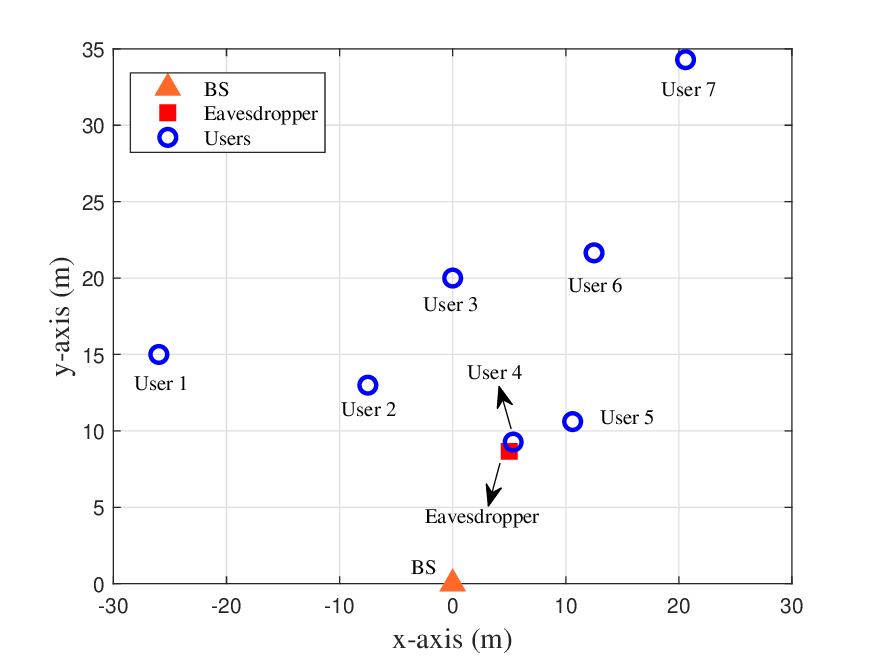}
\vspace*{-4mm}
\caption{Simulation setup of a sensing-assisted near-field secure communication system with $K=7$ users, one eavesdropper and one BS.}
\label{figure:exp_illus}
\end{figure}

In this section, we use simulation results to validate the effectiveness of the proposed algorithms. As shown in Fig. \ref{figure:exp_illus}, we consider a system with one BS, $K=7$ users, and one eavesdropper which moves around to wiretap the users' information. 
The BS is equipped with a ULA with an aperture of $1$ m \cite{9903389,9422343, 10579914} and the system operates at the frequency $f_c = 28$ GHz. As a result, the  Rayleigh distance is $186$ m.
The ULA consists of $N=64$ antennas and is centered at the origin of the coordinate system parallel to the x-axis. Unless otherwise specified, we adopt the following parameters $\Delta T = 0.2$ s, $\sigma_{\theta_{\mathrm{E}}} = 0.02^{\circ}$, $\sigma_{d_{\mathrm{E}}}=0.2$ m, $\sigma_{v_x} = 0.15$ m/s,  $\sigma_{v_y} = 0.15$ m/s \cite{9171304},
$a_{\tau} = 1 \times 10^{-6}$, $a_{\nu} = 600$, $a_{\theta_{E}} = 0.1$ \cite{10476610}, $P_{\mathrm{max}} = 37$ dBm, $\overline{R}_{\mathrm{info},k}=6$ bits/s/Hz, $\overline{R}_{\mathrm{leak},k}=0.05$ \cite{10345500}, $\sigma_{k}^2 = -70$ dBm, $\sigma_s^2 = -80$ dBm, $\sigma_e^2 = -80$ dBm and $G=10^4$ \cite{10476610}.
\par
For simplicity, we refer to the proposed GBD-based optimal design and the ZF-SCA-based low-complexity design as ``optimal" and ``low-complexity", respectively. For comparison purposes, we include the scheme that selects the served user based on the channel correlation coefficient between users and the eavesdropper \cite{10345500}, which is referred to as the ``correlation-based scheduling" in the figures. The channel correlation coefficient between the $k$-th user and the eavesdropper is defined by $\varpi_k =\left| \mathbf{h}_k^H \mathbf{h}_{\mathrm{E}} \right|/(\left\| \mathbf{h}_k \right\| \left\| \mathbf{h}_{\mathrm{E}} \right\|)$,
where $\mathbf{h}_{\mathrm{E}}$ is calculated based on the predicted location because the true value is unknown.
The system selects users with the first $\Gamma_1$ smallest correlation coefficients.

\begin{figure}[t]
\centering
\includegraphics[width=3.5in]{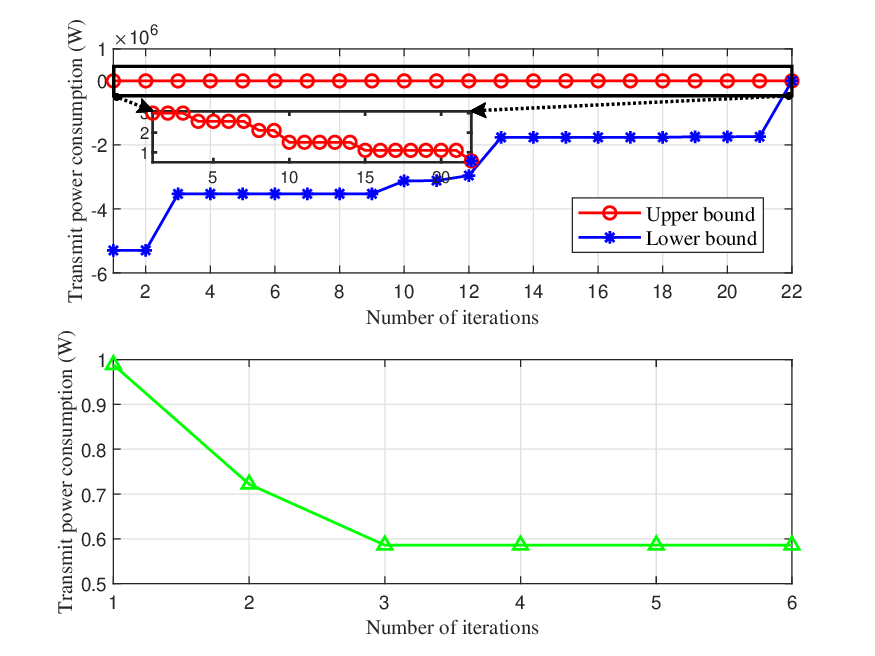}
\vspace*{-6mm}
\caption{Convergence of \textbf{Algorithm 1} (Top) and \textbf{Algorithm 2} (Bottom).}
\label{figure:Convergence}
\vspace*{0mm}
\end{figure}


\begin{figure}[t]
\centering
\includegraphics[width=3.4in]{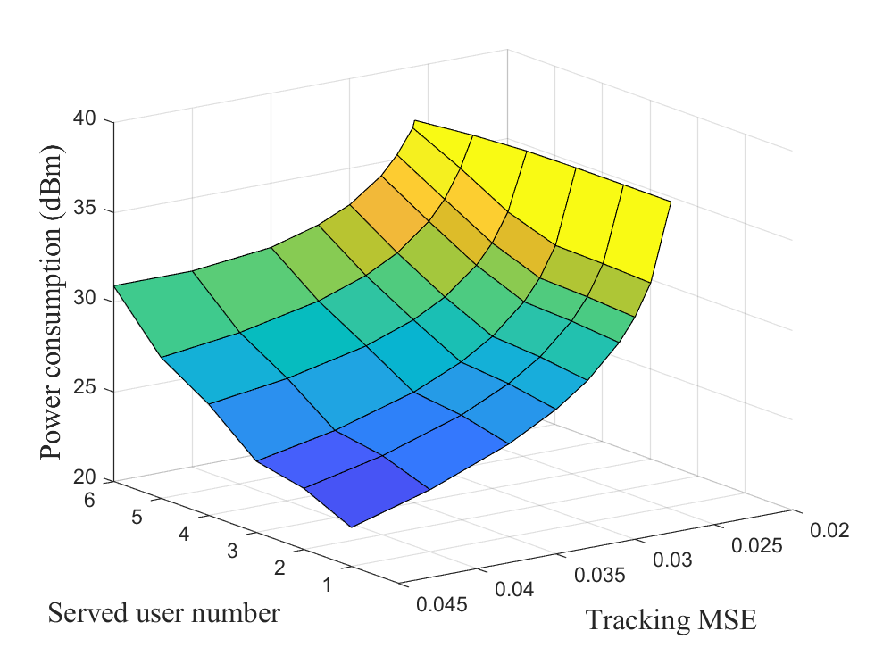}
\vspace*{-4mm}
\caption{Optimal Pareto boundary.}
\label{figure:pareto}
\end{figure}
\subsection{Algorithm Convergence}
Fig. \ref{figure:Convergence} presents the convergence behavior of the proposed algorithms, where $\Gamma_1$ and $\Gamma_2$ are set to $5$ and $0.1$, respectively. As shown in the figure, with \textbf{Algorithm 1}, the upper bound of the objective function value is non-increasing, the lower bound is non-decreasing, and they converge to the same point. This indicates that \textbf{Algorithm 1} will converge to the optimal solution, which validates the convergence analysis for the GBD-based optimal design in Section IV-B. In addition, it can be observed that \textbf{Algorithm 2} converges in several iterations.

\subsection{Optimal Pareto Boundary}
The trade-offs between the transmit power, tracking performance, and number of served users are depicted through the optimal Pareto boundary in Fig. \ref{figure:pareto}, which consists of the Pareto optimal solutions obtained by the proposed algorithm. Note that, with the proposed algorithm, we can quickly determine the edge of the Pareto boundary by relaxing the specific constraints of \eqref{constrained_moop}, and hence avoid the exhaustive search over the whole space. For the scenario considered in Fig. \ref{figure:exp_illus},
the maximum number of served users is 6 because the eavesdropper is very close to user $4$, such that user 4 can not be scheduled due to information leakage rate constraint.
As shown in Fig. \ref{figure:pareto}, optimizing any one performance metric leads to the degeneration of the other two. For example, to increase the number of served users, the system should either allocate more power or degenerate the tracking performance. 

\begin{figure}[t]
  \centering
      \subfigure[Sensing beampattern toward the eavesdropper located at $(63^{\circ}, 3 \, \text{m})$.]{
        \includegraphics[width=0.46\linewidth]{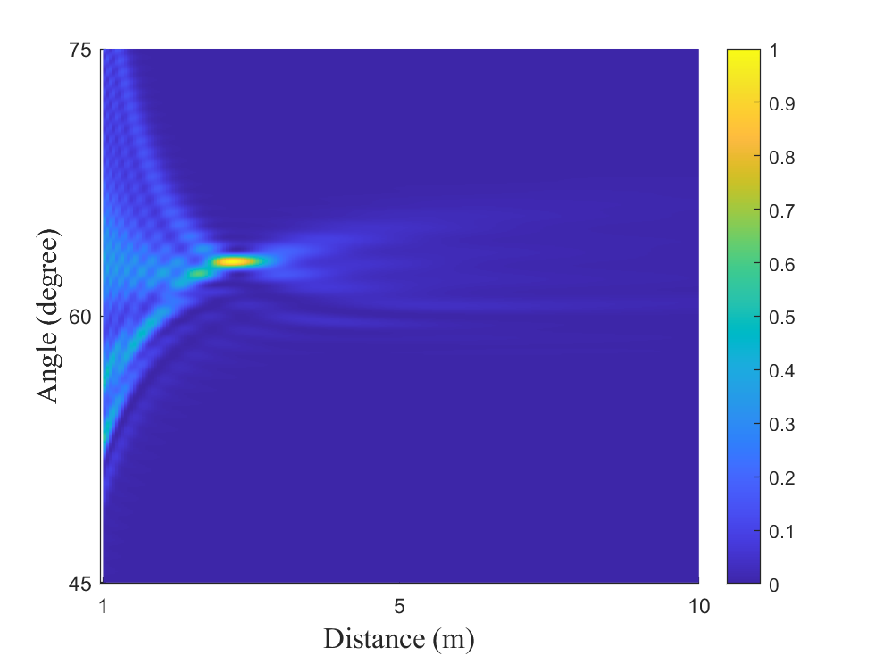}
        \label{figure:beampattern_a}
      }
      \subfigure[Information beampattern of the user. Eavesdropper is located at $(63^{\circ}, 3 \, \text{m}).$]{
        \includegraphics[width=0.46\linewidth]{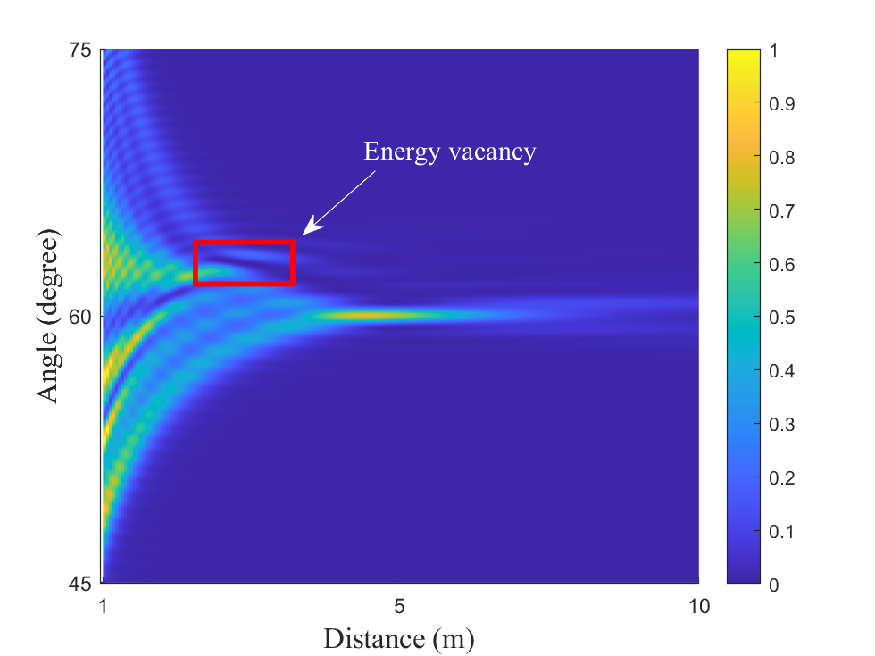}
        \label{figure:beampattern_b}
      }
      \subfigure[Sensing beampattern toward the eavesdropper located at $(60^{\circ}, 3 \, \text{m})$.]{
        \includegraphics[width=0.46\linewidth]{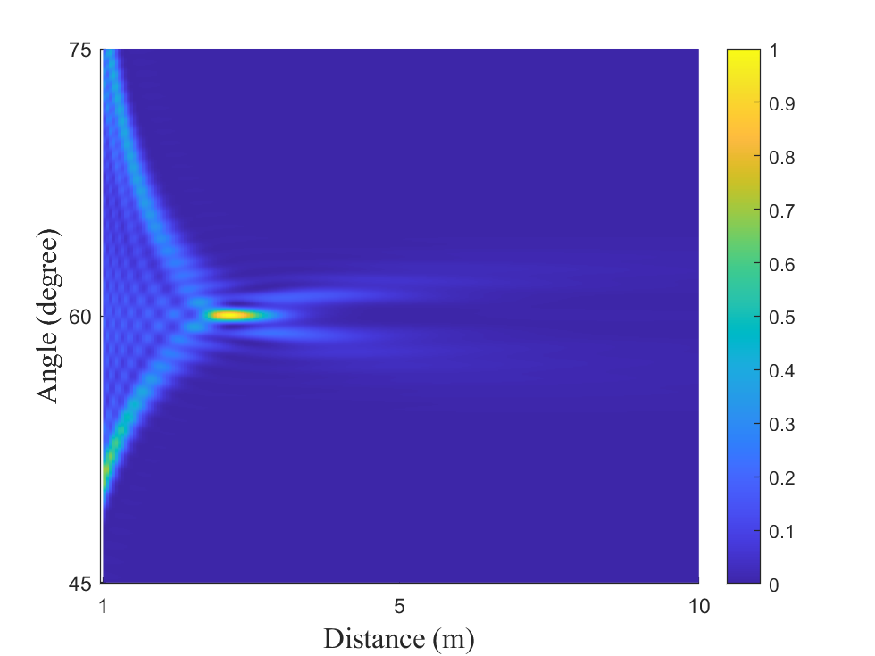}
        \label{figure:beampattern_c}
      }
      \subfigure[Information beampattern of the user. Eavesdropper is located at $(60^{\circ}, 3 \, \text{m}).$]{
        \includegraphics[width=0.46\linewidth]{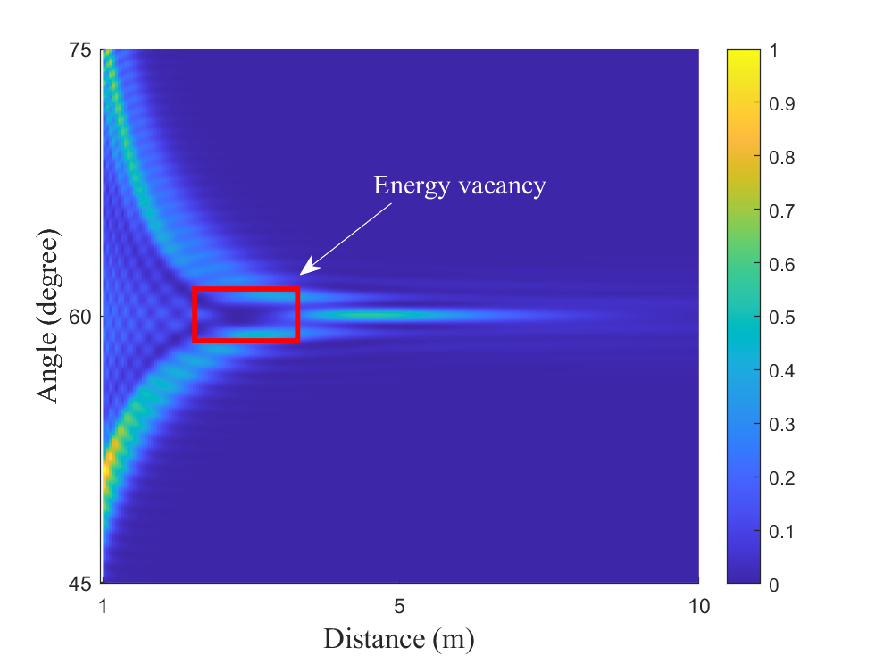}
        \label{figure:beampattern_d}
      }
    \caption{Normalized 2D beampattern of sensing-assisted near-field secure communication.}
    \label{figure:beampattern}
    \vspace*{-2mm}
  \end{figure}
\begin{figure}[t]
\centering
\includegraphics[width=3.2in]{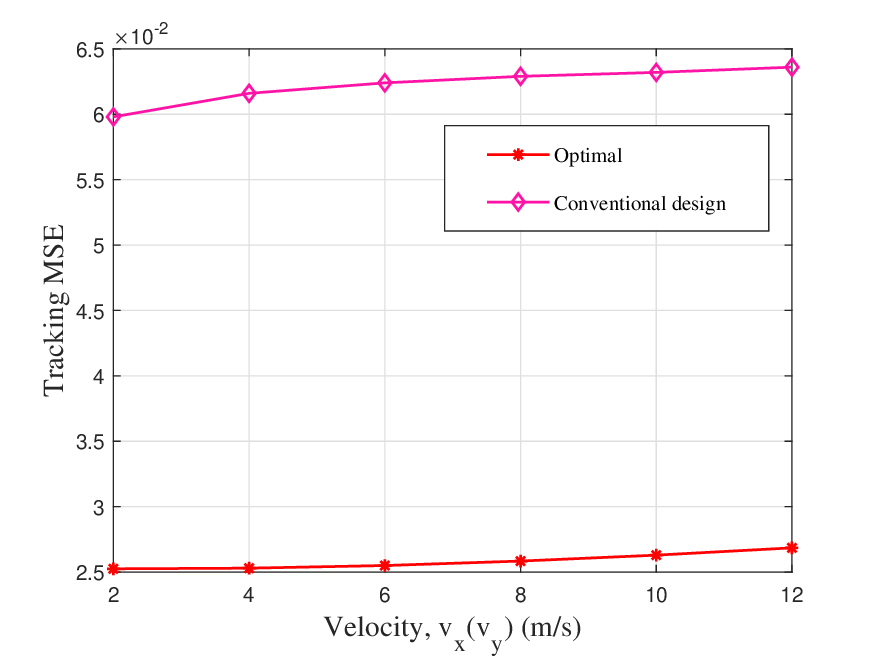}
\vspace*{-4mm}
\caption{Minimum tracking MSE versus velocity of eavesdropper.}
\label{figure:eps_velocity}
\vspace*{0mm}
\end{figure}
\subsection{Impact of Beam Focusing and Beam Diffraction in Near-field Secure Communication}
In this sub-section, we investigate the unique properties of the near-field beamforming and their influence on the PLS design. For ease of illustration, we consider one user located at $(60^{\circ}, 5.5 \, \text{m})$. When the eavesdropper is located at $(63^{\circ}, 3 \, \text{m})$, the beampatterm of the dedicated sensing signal and information signal are presented in Fig. \ref{figure:beampattern_a} and Fig. \ref{figure:beampattern_b}, respectively. It can be observed from Fig. \ref{figure:beampattern_a} that the dedicated sensing signal is focused at the eavesdropper for sensing and jamming purposes. On the one hand, Fig. \ref{figure:beampattern_b} shows that the information signal is focused on the user while bypassing the eavesdropper. This phenomenon is described as beam diffraction in \cite{liu2024ris}.
On the other hand, as shown in Fig. \ref{figure:beampattern_c} and \ref{figure:beampattern_d}, when the eavesdropper is located at $(60^{\circ}, 3 \, \text{m})$, i.e., with the same AoD as the user, the  BS can still perform secure transmission by exploiting the distance DoF. 
Note that there is no feasible solution for the above case in far-field communication.

\begin{figure}[t]
\centering
\includegraphics[width=3.2in]{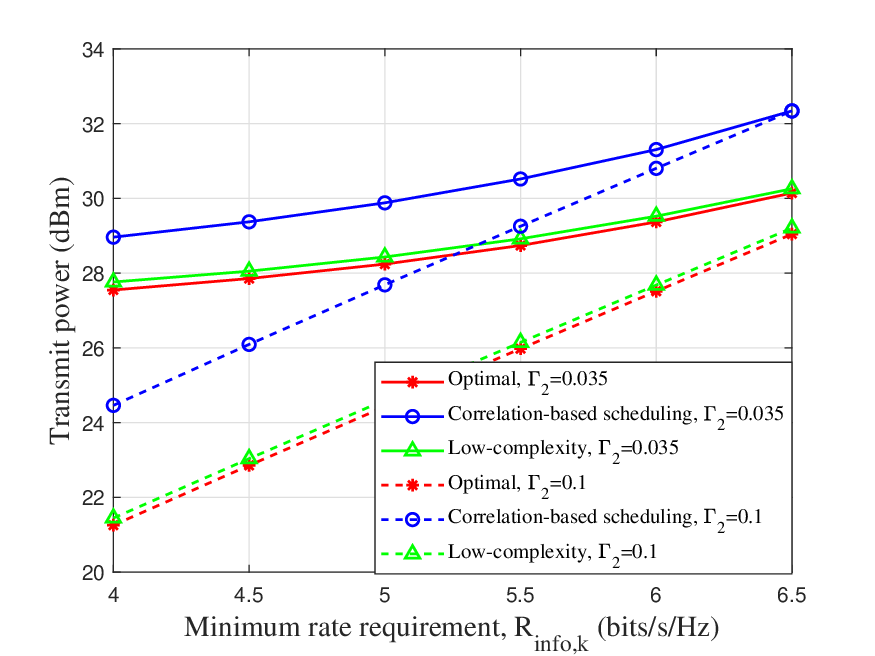}
\vspace*{-4mm}
\caption{Minimum transmit power versus achievable rate requirement, $\Gamma_1=5$.}
\label{figure:E_rate}
\end{figure}

\begin{figure}[t]
\centering
\includegraphics[width=3.2in]{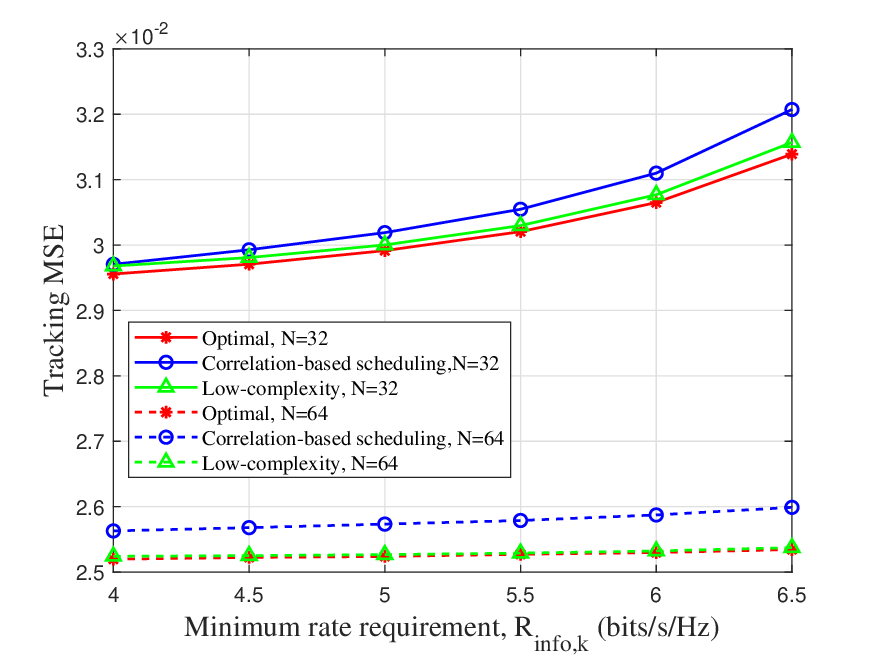}
\vspace*{-2mm}
\caption{Minimum tracking MSE versus achievable rate requirement, $\Gamma_1=5$.}
\label{figure:eps_rate}
\vspace*{0mm}
\end{figure}
\begin{figure}[t]
\centering
\includegraphics[width=3.2in]{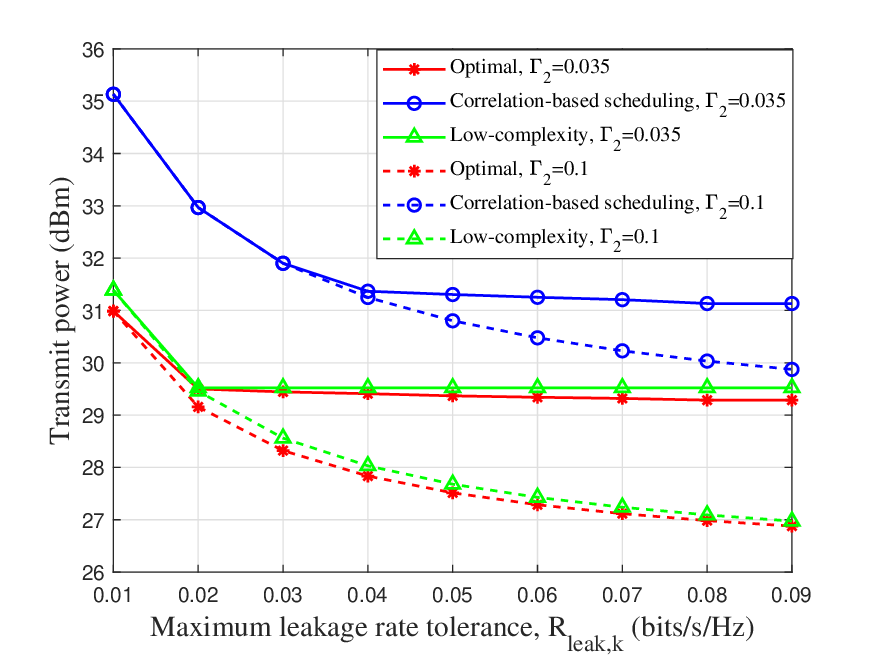}
\vspace*{-2mm}
\caption{Minimum transmit power versus maximum leakage rate tolerance, $\Gamma_1=5$.}
\label{figure:E_leak}
\vspace*{0mm}
\end{figure}


\begin{figure}[t]
\centering
\includegraphics[width=3.3in]{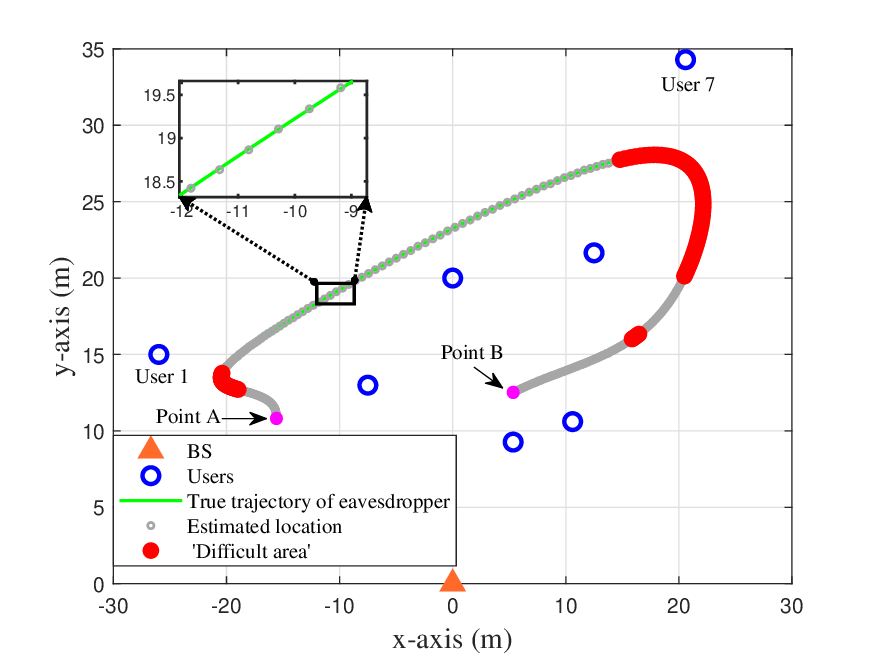}
\vspace*{-4mm}
\caption{Estimated trajectory of eavesdropper.}
\label{figure:trajectory}
\end{figure}
\begin{figure}[t]
\centering
\includegraphics[width=3.3in]{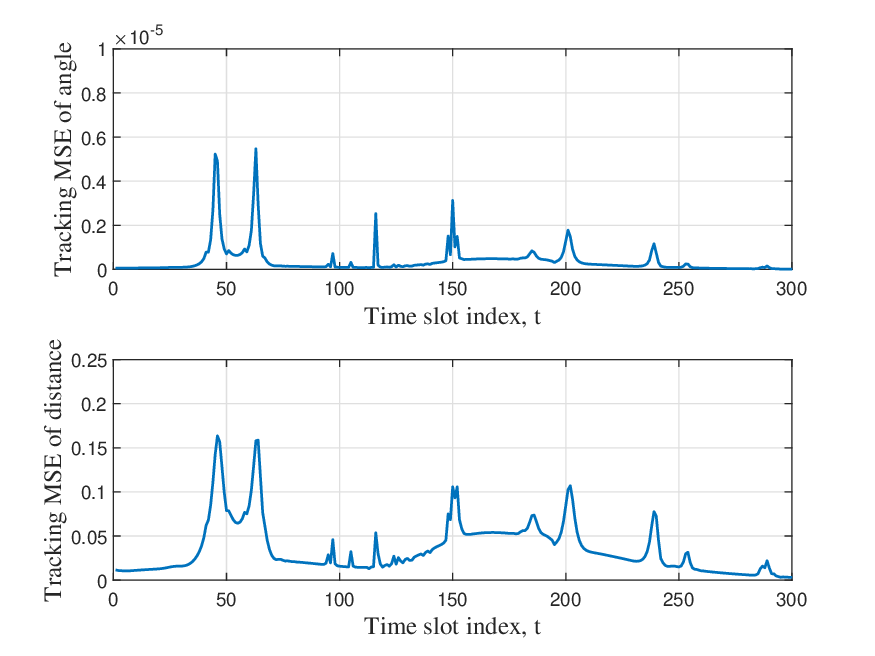}
\vspace*{-4mm}
\caption{Tracking performance.}
\label{figure:tracking_error}
\end{figure}
\par
\subsection{Effect of the Mobility of Eavesdropper}
We investigate the effect of the mobility of the eavesdropper on tracking performance in Fig. \ref{figure:eps_velocity}. It can be observed that the tracking MSE increases with the velocity. Note that even for the high mobility situations ($v_x=v_y=12$ m/s), the proposed algorithm still provides a satisfactory tracking performance. For comparison purposes, we also include the conventional design where the beamforming is based on the estimated location in the previous time slot.
It can be observed that the optimal design outperforms the conventional one by a significant gap due to the more accurate CSI provided by the tracking scheme.
\par
\subsection{Effect of the Achievable Rate Requirement}
We investigate the effect of the achievable rate requirement in Fig. \ref{figure:E_rate} and Fig. \ref{figure:eps_rate}, where the minimum number of served users $\Gamma_1$ is set to $5$. In Fig. \ref{figure:E_rate}, we minimize the transmit power with different tracking performance $\Gamma_2$. It can be observed that the minimum transmit power increases with the increase of achievable rate requirement and the decrease of tracking performance threshold. The optimal scheme achieves the minimum power consumption compared with the other schemes. The gap between the correlation-based scheduling comes from optimal user scheduling. In addition, the low-complexity scheme achieves a performance close to the optimal scheme, indicating the effectiveness of the joint user scheduling and power allocation design.
\par
Furthermore, it can be observed from Fig. \ref{figure:E_rate} that the power gap caused by different $\Gamma_2$ decreases when $R_{\mathrm{info},k}$ increases.  
This is because higher achievable rate requirement leads to an increased risk of information leakage and the system needs to allocate more energy to the dedicated sensing signal for information jamming.
For $\Gamma_2=0.1$, with the increasing achievable rate, the increment of transmit power comes from two aspects: the information beamforming and the dedicated sensing beamforming. For a smaller $\Gamma_2=0.035$, the strict tracking performance requirement on dedicated sensing signal has already guaranteed a low information leakage rate. Hence, the increment of transmit power mainly comes from the information beamforming. This indicates that a high achievable rate requirement intrinsically requires an accurate tracking performance. 
\par
Fig. \ref{figure:eps_rate} shows the minimum tracking MSE versus the achievable rate requirement $\overline{R}_{\mathrm{info},k}$ with different numbers of antennas. The tracking MSE increases with $\overline{R}_{\mathrm{info},k}$, i.e., the improvement of quality-of-service comes at the cost of radar performance, implying the trade-off between communication and radar performance. In addition, it can be observed that a larger number of antennas brings more accurate tracking performance due to two reasons. On the one hand, more antennas provide more DoFs and enlarge the feasible set, leading to a better objective value. On the other hand, more antennas can achieve a more flexible and sophisticated beampattern for system operation.
\par
\subsection{Effect of the Maximum Leakage Rate Tolerance Factor}
We explore the effect of the maximum leakage rate tolerance factor in Fig. \ref{figure:E_leak}, where $\Gamma_1$ is set to $5$.
As shown in the figure, 
as $\overline{R}_{\mathrm{leak},k}$ increases, the transmit power first decreases and then remains almost unchanged under the stringent tracking requirement. This indicates that the stringent tracking requirement will naturally lead to a small information leakage rate due to the fact that the sensing signal is also utilized for jamming purposes. Take the optimal design for instance, the dedicated sensing signal satisfying $\Gamma_2=0.035$ leads to an information leakage rate around $0.02$ bits/s/Hz.
\par
\subsection{Tracking Eavesdropper via Low-complexity Design}
In this subsection, we aim to validate the performance of the low-complexity algorithm by a practical use case. In particular, we choose the Pareto optimal point where the system achieves the minimum tracking error with the maximum number of users.
As shown in Fig. \ref{figure:trajectory}, the eavesdropper moves from point A to point B as indicated by the green real line. The estimated location of the eavesdropper is shown in the gray circle, and the tracking MSE of the angle and the distance are presented in Fig. \ref{figure:tracking_error}. It can be observed that the low-complexity algorithm provides satisfactory performance.
\par
Moreover, the regions of the trajectory with higher tracking errors, i.e., the regions where the tracking error for distance is larger than $0.05$, are highlighted and labeled as the ``difficult area" in Fig. \ref{figure:trajectory}. For the left ``difficult area'', the eavesdropper is between user $1$ and the BS. For the right ``difficult area'', the eavesdropper appears between user $7$ and the BS with multiple users in a similar direction, which is intractable for the far-field transmission.
\section{Conclusion}
This paper studied sensing-aided PLS in near-field communication systems with mobile eavesdroppers. 
To obtain a thorough analysis, a Pareto optimization framework was proposed to investigate the fundamental trade-offs of three critical system performance metrics: power consumption, number of securely served users, and tracking error. The optimal Pareto boundary was characterized by optimally solving the resulting MOOP. For that purpose, we transformed the considered MINLP into a tractable form and then developed an optimal algorithm based on GBD.
Then, a low-complexity ZF-based algorithm was developed, where the binary constraints were handled by the penalty method, and SCA was utilized to convexify the penalty term.
Simulation results validated the effectiveness of the proposed algorithms in dealing with mobile eavesdroppers.
In addition, it was observed that with the additional DoF in the distance domain, near-field PLS presents a beam diffraction effect in which the energy of the information beam is nulled around the eavesdropper and focused on the users. This enables near-field communication to cope with the situation where the eavesdropper is located in the same direction as the user.


\bibliographystyle{IEEEtran}
\bibliography{Reference_List}
\end{document}